\documentclass[a4paper, 11pt]{article}
\pdfoutput=1

\usepackage{jheppub}
\usepackage{afterpage}
\usepackage{multirow}
\usepackage{rotating}

\newcommand{\Dmq}{\Delta m^2}
\newcommand{\dCP}{\delta_\mathrm{CP}}
\newcommand{\Nuc}[2]{{\ensuremath{\mbox{}^{#1}}\text{#2}}}
\newcommand{\eVq}{\ensuremath{\text{eV}^2}}

\renewcommand{\Im}{\mathop{\mathrm{Im}}}
\newenvironment{pagefigure}{\begin{figure}[!p]}{\afterpage\clearpage\end{figure}}

\title{The fate of hints: updated global analysis of three-flavor
  neutrino oscillations}

\author[a]{Ivan Esteban,}
\affiliation[a]{Departament de Fis\'{\i}ca Qu\`antica i
  Astrof\'{\i}sica and Institut de Ciencies del Cosmos, Universitat de
  Barcelona, Diagonal 647, E-08028 Barcelona, Spain}
\emailAdd{ivan.esteban@fqa.ub.edu}

\author[a,b,c]{M.~C.~Gonzalez-Garcia,}
\affiliation[b]{Instituci\'o Catalana de Recerca i Estudis
  Avan\c{c}ats (ICREA), Pg. Lluis Companys 23, 08010 Barcelona,
  Spain.}
\affiliation[c]{C.N.~Yang Institute for Theoretical Physics, State
  University of New York at Stony Brook, Stony Brook, NY 11794-3840,
  USA}
\emailAdd{maria.gonzalez-garcia@stonybrook.edu}

\author[d]{Michele Maltoni,}
\affiliation[d]{Instituto de F\'{\i}sica Te\'orica UAM/CSIC, Calle de
  Nicol\'as Cabrera 13--15, Universidad Aut\'onoma de Madrid,
  Cantoblanco, E-28049 Madrid, Spain}
\emailAdd{michele.maltoni@csic.es}

\author[e]{Thomas Schwetz,}
\affiliation[e]{ Institut f\"ur Kernphysik, Karlsruher Institut f\"ur
  Technologie (KIT), D-76021 Karlsruhe, Germany}
\emailAdd{schwetz@kit.edu}

\author[e]{Albert Zhou}
\emailAdd{albert.zhou@kit.edu}

\abstract{Our herein described combined analysis of the latest neutrino
  oscillation data presented at the Neutrino2020 conference shows that
  previous hints
  for the neutrino mass ordering have significantly decreased, and
  normal ordering (NO) is favored only at the $1.6\sigma$
  level. Combined with the $\chi^2$ map provided by Super-Kamiokande
  for their atmospheric neutrino data analysis the hint for NO is at
  $2.7\sigma$.  The CP conserving value $\dCP = 180^\circ$ is within
  $0.6\sigma$ of the global best fit point. Only if we restrict to
  inverted mass ordering, CP violation is favored at the $\sim
  3\sigma$ level.
  We discuss the origin of these results -- which are
  driven by the new data from the T2K and NOvA long-baseline
  experiments--,  and the relevance of the LBL-reactor oscillation frequency
  complementarity.
  The previous $2.2\sigma$ tension in $\Dmq_{21}$
  preferred by KamLAND and solar experiments is also reduced to the
  $1.1\sigma$ level after the inclusion of the latest Super-Kamiokande
  solar neutrino results. Finally we present updated allowed ranges
  for the oscillation parameters and for the leptonic Jarlskog
  determinant from the global analysis.}

\preprint{IFT-UAM/CSIC-112, YITP-SB-2020-21}

\keywords{neutrino oscillations, solar and atmospheric neutrinos}

\begin{document}

\maketitle

\section{Introduction}

Global fits to neutrino oscillation data in the last several years
have shown persistent hints for the normal neutrino mass ordering and
values of the CP phase $\dCP$ around maximal CP
violation~\cite{Esteban:2016qun, Esteban:2018azc, deSalas:2020pgw,
  deSalas:2018bym, Capozzi:2020qhw, Capozzi:2018ubv}. In this article
we are going to re-assess the status of those hints in light of the
new data released at the Neutrino2020 conference, in particular by the
T2K~\cite{Abe:2019vii, T2K:nu2020} and NOvA~\cite{Acero:2019ksn,
  NOvA:nu2020} long-baseline (LBL) experiments. As we are going to
discuss in detail, the hints have mostly disappeared or are
significantly decreased: both neutrino mass orderings provide fits of
comparable quality to the global data from accelerator and reactor
experiments, and the CP conserving value $\dCP = 180^\circ$ is within
the $1\sigma$ allowed range.

We discuss in detail the origin of this apparent change of trends and
trace back the data samples responsible for the change. We are going
to compare the latest status with our pre-Neutrino2020 analysis,
NuFIT 4.1, available at the NuFIT website~\cite{nufit}. Most relevant
for mass ordering and CP phase are the updates of the neutrino samples
for T2K~\cite{T2K:nu2020}, from $1.49$ to $1.97\times 10^{21}$~POT,
and NOvA~\cite{NOvA:nu2020}, from $0.885$ to $1.36\times 10^{21}$~POT.
The T2K and NOvA anti-neutrino exposures are the same as used for
NuFIT 4.1, but both collaborations introduced relevant changes in
their analysis and hence we have adapted also our anti-neutrino fits
correspondingly.
In addition we have updated the reactor experiments
Double-Chooz~\cite{DoubleChooz:2019qbj, DoubleC:nu2020} from 818/258 to 1276/587 days of
far/near detector data and RENO~\cite{Bak:2018ydk, RENO:nu2020} from 2200 to 2908
days of exposure.

Another update concerns the solar neutrino oscillation analysis, to
include the latest total energy spectrum and the day-night asymmetry
of the SK4 2970-day sample presented at
Neutrino2020~\cite{SK:nu2020}. As we will show, thanks to these new
data the tension on the determination of $\Dmq_{21}$ from KamLAND
versus solar experiments has basically disappeared.

The outline of the paper is as follows. In Sec.~\ref{sec:lbl} we
discuss the status of the neutrino mass ordering and the leptonic CP
phase $\dCP$, focusing on recent updates from T2K, NOvA, as well as
the combination of LBL accelerator and reactor experiments. Despite
somewhat different tendencies, we will show quantitatively that
results from T2K and NOvA as well as reactors are fully statistically
compatible.  The status of the tension between solar and KamLAND
results is presented in Sec.~\ref{sec:sol-kam}.
Section~\ref{sec:globalsum} contains a selection of the combined
results of this global fit, NuFIT 5.0, which updates our previous
analyses~\cite{GonzalezGarcia:2012sz, Gonzalez-Garcia:2014bfa,
  Esteban:2016qun, Esteban:2018azc}. In particular we present the
ranges of allowed values for the oscillation parameters and
of the leptonic Jarlskog determinant.\footnote{Additional figures,
  $\Delta\chi^2$ maps and future updates of this analysis will be made
  available at the NuFIT website~\cite{nufit}.}  Parametrization
conventions and technical details on our global analysis can be found
in Ref.~\cite{Esteban:2018azc}.  In particular, in what follows we use
the definition
\begin{equation}
  \Dmq_{3\ell}
  \quad \text{with}\quad
  \begin{cases}
    \ell = 1 & \text{for $\Dmq_{3\ell} > 0$: normal ordering (NO),} \\ 
    \ell = 2 & \text{for $\Dmq_{3\ell} < 0$: inverted ordering (IO).}
  \end{cases}
\end{equation}
We finish by summarizing our results in Sec.~\ref{sec:conclu}.  A full
list of the data used in this analysis is given in
appendix~\ref{sec:appendix-data}.

\section{Fading hints for CP violation and neutrino mass ordering}
\label{sec:lbl}

\subsection{T2K and NOvA updates}

We start by discussing the implications of the latest data from the
T2K and NOvA long-baseline accelerator experiments, presented at the
Neutrino2020 conference.\footnote{During the preparation of this work
  Ref.~\cite{Kelly:2020fkv} appeared presenting related partial
  results in qualitative agreement with some of our findings for the
  LBL analysis.}  To obtain a qualitative understanding we follow
Refs.~\cite{Elevant:2015ska, Esteban:2018azc} and expand the
oscillation probability relevant for the T2K and NOvA appearance
channels in the small parameters $\sin\theta_{13}$,
$\Dmq_{21}L/E_\nu$, and $A \equiv | 2E_\nu V / \Dmq_{3\ell}|$, where
$L$ is the baseline, $E_\nu$ the neutrino energy and $V$ the effective
matter potential~\cite{Wolfenstein:1977ue}:
\begin{align}
  P_{\nu_\mu\to\nu_e} & \approx 4 s_{13}^2s_{23}^2(1+2oA) - C \sin\dCP(1+oA) \,,
  \\
  P_{\bar\nu_\mu\to\bar\nu_e} &\approx 4 s_{13}^2s_{23}^2(1-2oA) + C \sin\dCP(1-oA) \,.
\end{align}
with $s_{ij} \equiv \sin\theta_{ij}$ and
\begin{equation}
  C \equiv \frac{\Dmq_{21}L}{4E_\nu}
  \sin2\theta_{12}\sin2\theta_{13}\sin2\theta_{23}\,,
  \quad
  o \equiv \text{sgn}(\Dmq_{3\ell}) \,,
\end{equation}
and we have used $|\Dmq_{3\ell}|\, L/4E_\nu \approx \pi/2$ for T2K and
NOvA.  At T2K, the mean neutrino energy gives $A \approx 0.05$,
whereas for NOvA we find that with the \emph{empirical} value of
$A=0.1$ the approximation works best. Assuming that the number of
observed appearance events in T2K and NOvA is approximately
proportional to the oscillation probability we obtain
\begin{align}
  N_{\nu_e}
  &\approx \mathcal{N}_\nu
  \left[ 2 s_{23}^2(1+2oA) - C' \sin\dCP(1+oA) \right] \,,
  \label{eq:Nnu}
  \\
  N_{\bar\nu_e} &
  \approx \mathcal{N}_{\bar\nu}
  \left[ 2 s_{23}^2(1-2oA) + C' \sin\dCP(1-oA) \right] \,.
  \label{eq:Nan}
\end{align}
Taking all the well-determined parameters $\theta_{13}$,
$\theta_{12}$, $\Dmq_{21}$, $|\Dmq_{3\ell}|$ at their global best fit
points, we obtain numerically $C' \approx 0.28$ with negligible
dependence on $\theta_{23}$. The normalization constants
$\mathcal{N}_{\nu,\bar\nu}$ calculated from our re-analysis of T2K and
NOvA are given for the various appearance samples in
table~\ref{tab:app}. Hence the expression in eqs.~\eqref{eq:Nnu}
and~\eqref{eq:Nan} serve well to understand the main behaviour under
varying the parameters $\sin^2\theta_{23}$, $\dCP$, and the mass
ordering.

In table~\ref{tab:app} we also show the observed number of events,
background subtracted events, as well as the ratio $r =
(N_\text{obs}-N_\text{bck}) / \mathcal{N}_{\nu(\bar\nu)}$. In a fit,
the values of $r$ have to be accommodated by the expression in the
square brackets of eqs.~\eqref{eq:Nnu} and~\eqref{eq:Nan}. In
brackets, we give also the $r$ values for the NuFIT 4.1 data set, to
illustrate the impact of the latest data.

\begin{table}\centering
  \begin{tabular}{c|cccccc}
    \hline\hline
    & \multicolumn{3}{c}{T2K ($\nu$)} & T2K ($\bar\nu$) & NOvA ($\nu$) & NOvA ($\bar\nu$) \\
    & CCQE & CC1$\pi$ & sum & \\
    \hline
    $\mathcal{N}$            & 48.5   & 5 & 53.5  & 16  & 48.5   & 23  \\
    $N_\text{obs}$             & 94   & 14 & 108  & 16   & 82   & 33 \\
    $N_\text{obs}-N_\text{bck}$  & 76 & 12.1 & 88.1  & 9.8 & 55.2 & 19 \\
    ratio $r$                & 1.6 (1.5) & 2.4 (3.6) & 1.65 (1.71) & 0.61 (0.7) & 1.14 (1.3) & 0.83 (0.7)\\
    \hline\hline
  \end{tabular}
  \caption{Normalization coefficients $\mathcal{N}_\nu$ and
    $\mathcal{N}_{\bar\nu}$ in eqs.~\eqref{eq:Nnu} and~\eqref{eq:Nan}
    for approximations used to qualitatively describe the appearance
    event samples for T2K and NOvA.  We also give the observed number
    of events, as well as the corresponding background subtracted
    event numbers, as reported in Refs.~\cite{T2K:nu2020,
      NOvA:nu2020}. The ratio in the last line is defined as $r =
    (N_\text{obs}-N_\text{bck}) / \mathcal{N}$ and numbers in brackets
    are the corresponding values for the data set used for NuFIT 4.1.}
  \label{tab:app}
\end{table}

\begin{figure}\centering
  \includegraphics[width=0.85\textwidth]{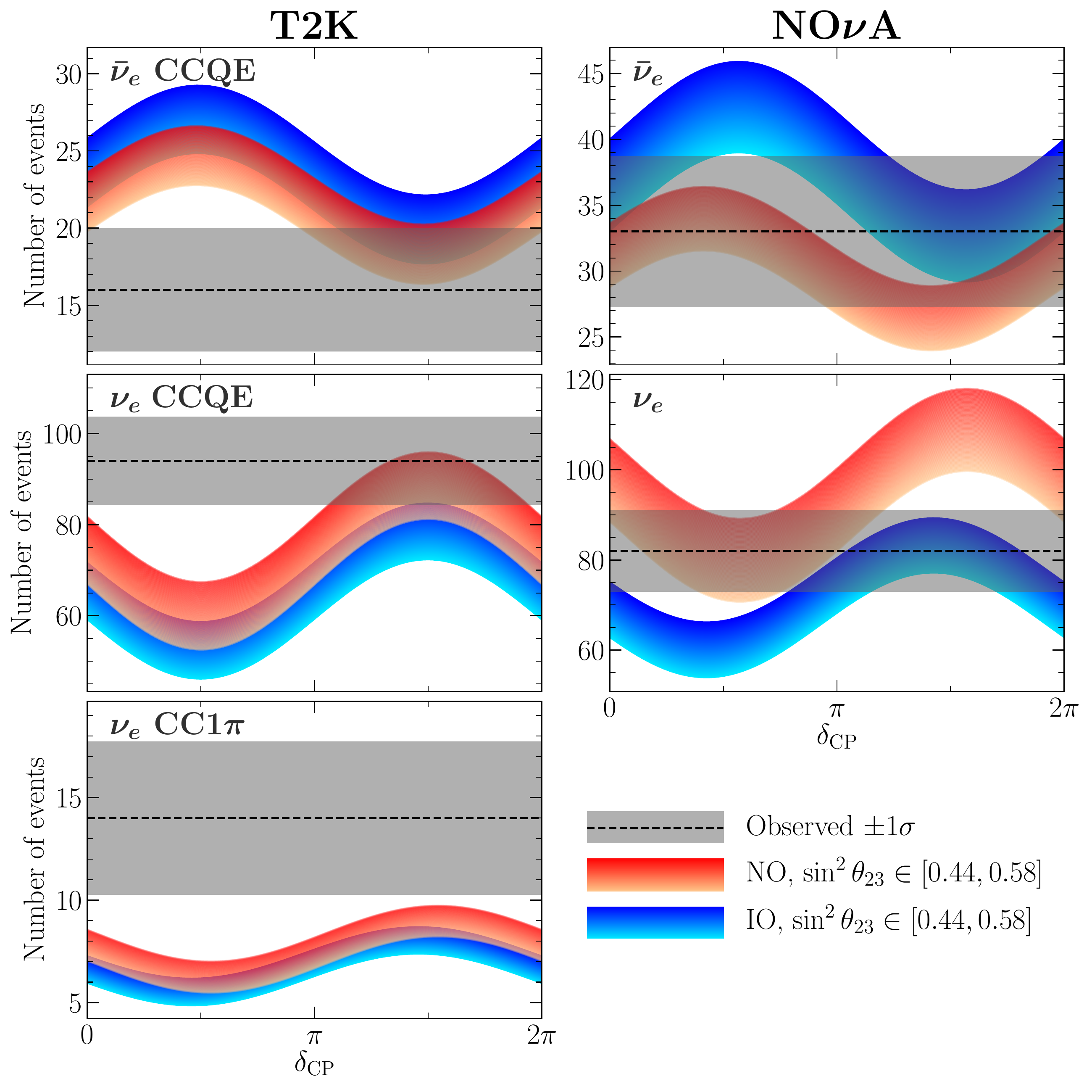}
  \caption{Predicted number of events as a function of $\dCP$ for the
    T2K (left) and NOvA (right) appearance data sets.
    $\sin^2\theta_{23}$ varies between 0.44 and 0.58, where the
    lower-light (upper-dark) bound of the colored bands corresponds to
    0.44 (0.58). Red (blue) bands correspond to NO (IO). For the other
    oscillation parameters we have adopted $\sin^2\theta_{13} =
    0.0224$, $|\Dmq_{3\ell}| = 2.5\times 10^{-3}~\eVq$,
    $\sin^2\theta_{12} = 0.310$, $\Dmq_{21} = 7.39\times
    10^{-5}~\eVq$. The horizontal dashed lines show the observed
    number of events, with the $\pm 1\sigma$ statistical error
    indicated by the gray shaded band.}
  \label{fig:nevts}
\end{figure}

Similar information is presented graphically in
figure~\ref{fig:nevts}, showing the predicted number of events for the
various appearance event samples as a function of $\dCP$, changing
$\sin^2\theta_{23}$ as well as the ordering, compared to the observed
event number. Here the predictions are calculated using our experiment
simulation based on fully numerical oscillation probabilities, while
the general behaviour of the curves is well described by
eqs.~\eqref{eq:Nnu} and~\eqref{eq:Nan}.

We can clearly observe a number of tendencies. T2K data has $r > 1$
for neutrinos and $r < 1$ for anti-neutrinos, implying that the
square-bracket in~\eqref{eq:Nnu} [\eqref{eq:Nan}] has to be enhanced
[suppressed]. If $\theta_{13}$ is fixed as determined by reactor
experiments this can be achieved by choosing NO and $\dCP \simeq
3\pi/2$ (see Sec.~\ref{sec:accrea} for a consistent combination of
reactor and LBL data).  This has been the driving factor for previous
hints for NO and maximal CP violation. We observe from the last row in
table~\ref{tab:app} that indeed this tendency has become somewhat
weaker with the new data, though still clearly present. In this
respect an interesting role is played by the CC$1\pi$ event sample. A
value $r=3.6$ for NuFIT 4.1 shows a large excess of events in this
sample, which has come down to $r=2.4$ with the latest
data. Figure~\ref{fig:nevts} still shows, that even the most favorable
parameter choice cannot accomodate the observed number of events
within $1\sigma$.  It seems that part of previous hints can be
attributed to a statistical fluctuation in this sub-leading event
sample. Let us stress, however, that due to the small CC$1\pi$ event
numbers, statistical uncertainties are large. Indeed, CCQE neutrino
and anti-neutrino events consistently point in the same direction and
they are both fitted best with NO and maximal CP phase.

\begin{figure}\centering
  \includegraphics[width=0.7\textwidth]{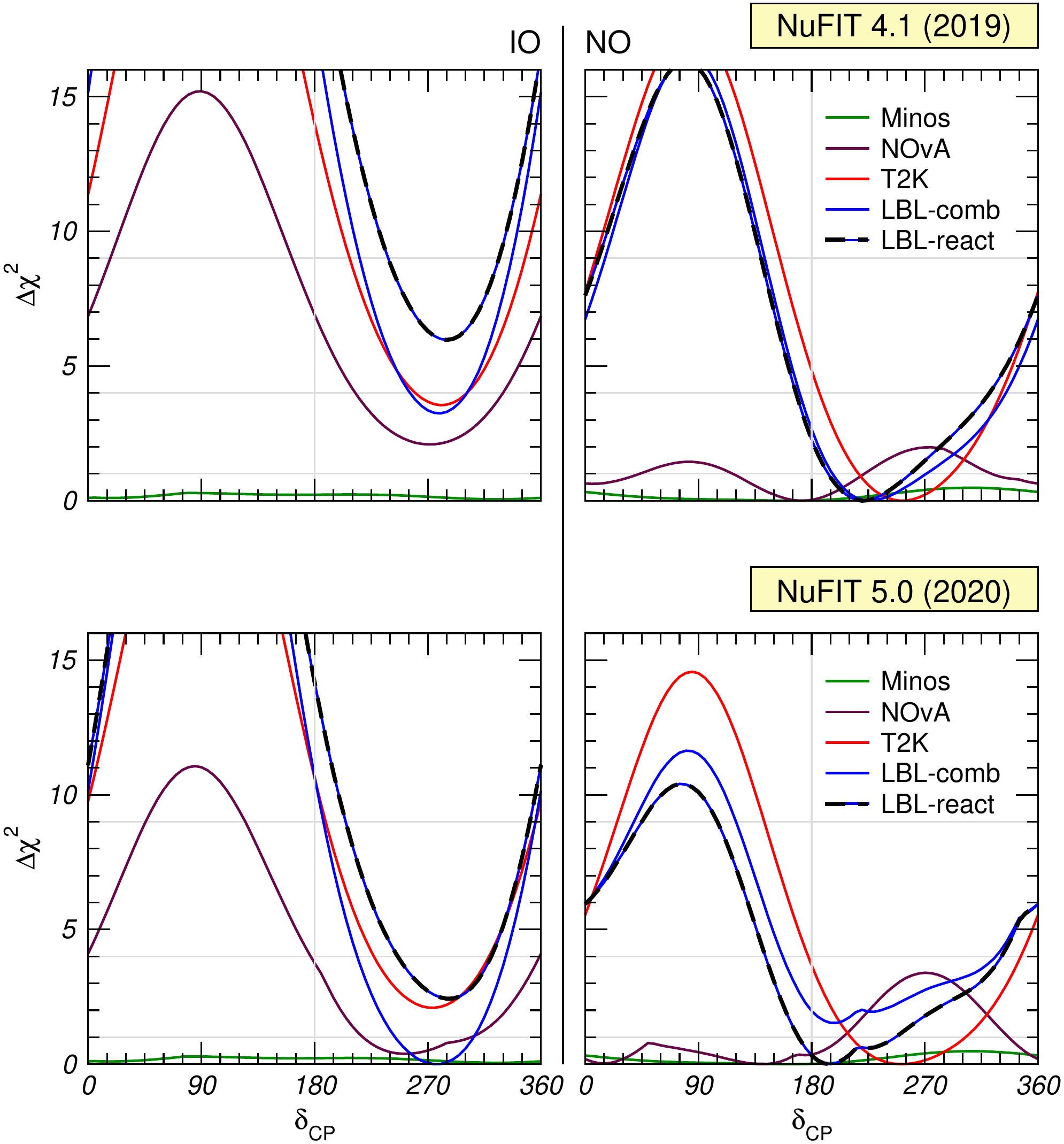}
  \caption{$\Delta\chi^2$ profiles as a function of $\dCP$ for
    different LBL data sets and their combination.  We have fixed
    $\sin^2\theta_{13}=0.0224$ as well as the solar parameters and
    minimized with respect to $\theta_{23}$ and $|\Dmq_{3\ell}|$. The
    black/blue dashed curves correspond to the combination of LBL data
    with the reactor experiments Day-aBay, RENO, Double-Chooz, and in
    this case also $\theta_{13}$ is left free in the fit. Left
    (right) panels are for IO (NO) and $\Delta\chi^2$ is shown with
    respect to the global best fit point for each curve. Upper panels
    are for the NuFIT 4.1 data set, whereas lower panels correspond to
    the current update.}
  \label{fig:compare-dcp}
\end{figure}

Moving now to NOvA, we first observe from figure~\ref{fig:nevts} the
larger separation between the NO and IO bands compared to T2K. This is
a manifestation of the increased matter effect because of the longer
baseline in NOvA. Next, neutrino data have $r\approx 1$ which can be
accommodated by (NO, $\dCP \simeq \pi/2$) or (IO, $\dCP \simeq
3\pi/2$). This behavior is consistent with NOvA anti-neutrinos,
however in tension with T2K in the case of NO. We conclude from these
considerations that the T2K and NOvA combination can be best fitted by
IO and $\dCP\simeq 3\pi/2$. This is indeed confirmed in
figure~\ref{fig:compare-dcp}, showing the $\Delta\chi^2$ profiles as a
function of $\dCP$. We observe in the lower-right panel that NOvA
disfavors (NO, $\dCP\simeq 3\pi/2$) by about 4 units in $\chi^2$,
whereas in the lower-left panel we see for IO consistent preference of
T2K and NOvA for $\dCP\simeq 3\pi/2$. For the combination this leads
to a preferred best fit for IO with $\Delta\chi^2(\text{NO}) \approx
1.5$ (which of course is not significant). We can also see that this
effect was less relevant in NuFIT 4.1 (fig.~\ref{fig:compare-dcp},
upper panels) for which we had $r=1.3$ --~compared to current
$1.14$~-- for NOvA neutrino data. This slightly higher ratio allowed
some more enhancement of the square-bracket in eq.~\eqref{eq:Nnu}
compared to the present situation, leading to less tension between T2K
and NOvA for NO. It also lead to a larger significance of NOvA for NO.

\begin{figure}\centering
  \includegraphics[width=\textwidth]{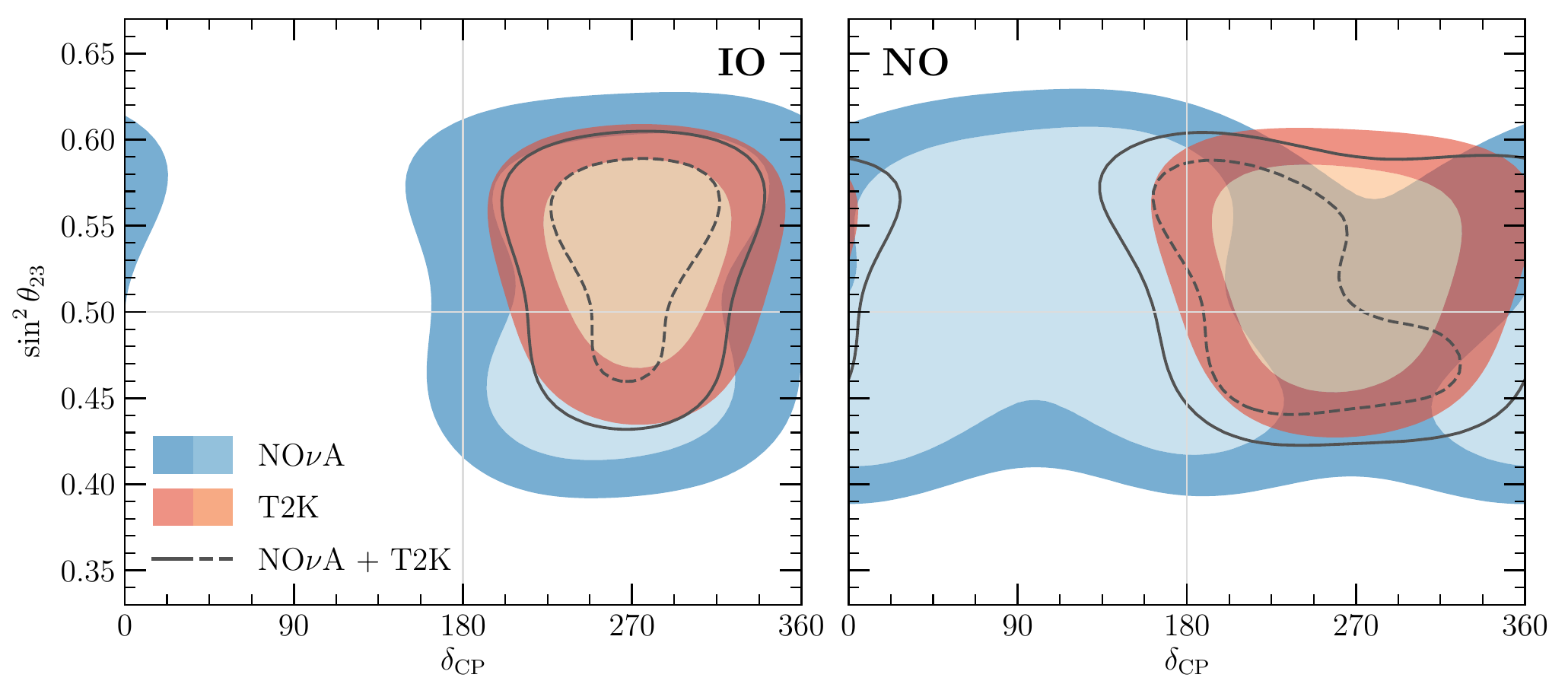}
  \caption{$1\sigma$ and $2\sigma$ allowed regions (2 dof) for T2K
    (red shading), NOvA (blue shading) and their combination (black
    curves). Contours are defined with respect to the local minimum
    for IO (left) or NO (right). We are fixing
    $\sin^2\theta_{13}=0.0224$, $\sin^2\theta_{12}=0.310$,
    $\Dmq_{21}=7.40\times 10^{-5}~\eVq$ and minimize with respect to
    $|\Dmq_{3\ell}|$.}
  \label{fig:sq23-dCP}
\end{figure}

The two-dimensional regions for T2K and NOvA in the ($\dCP$,
$\sin^2\theta_{23}$) plane for fixed $\theta_{13}$ are shown in
figure~\ref{fig:sq23-dCP}. The better consistency for IO is apparent,
while we stress that even for NO the $1\sigma$ regions touch each
other, indicating that also in this case the two experiments are
statistically consistent. We are going to quantify this later in
section~\ref{sec:PG}.

\subsection{Accelerator versus reactor}
\label{sec:accrea}

\begin{figure}\centering
  \includegraphics[width=0.48\textwidth]{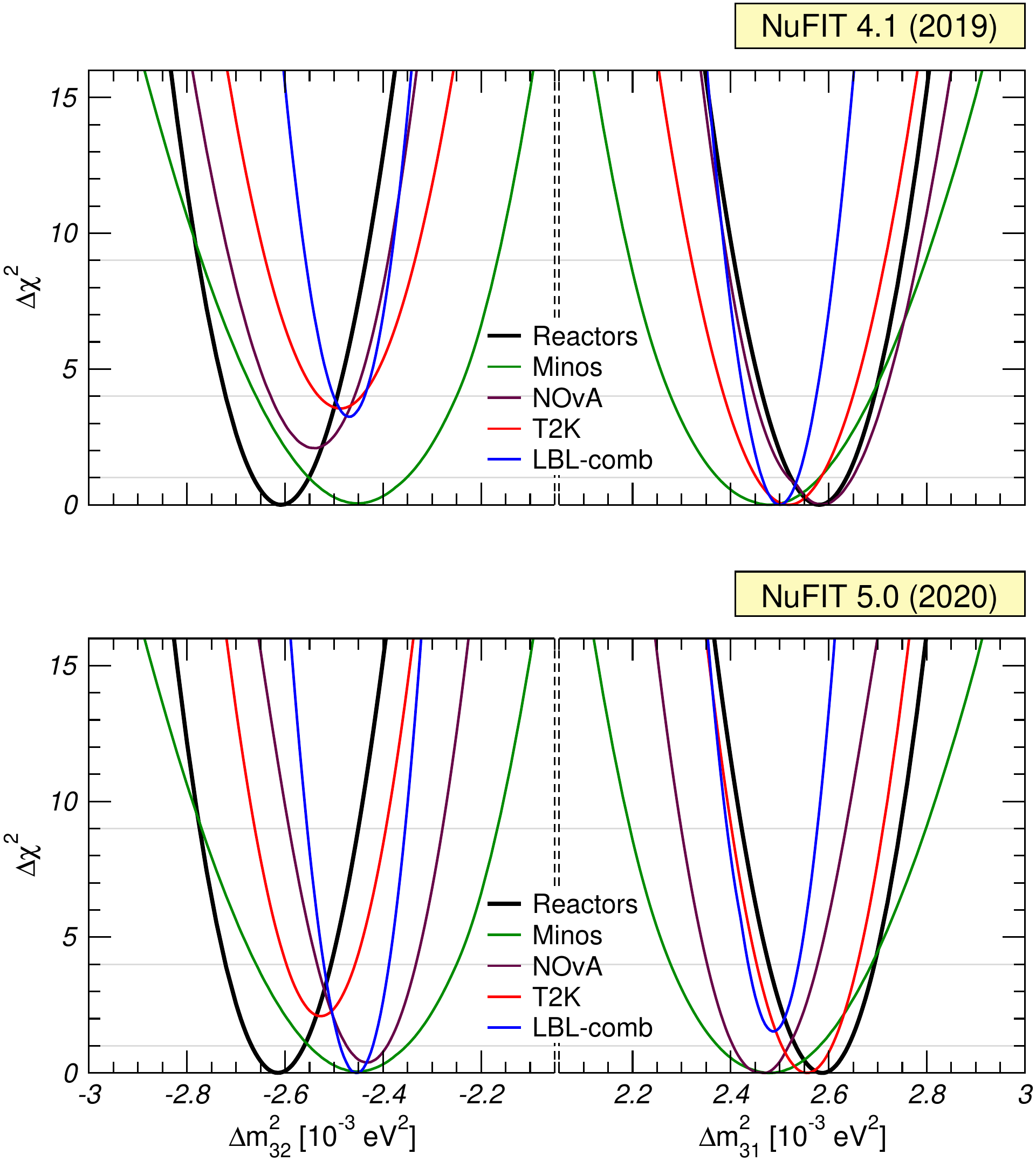}
  \includegraphics[width=0.49\textwidth]{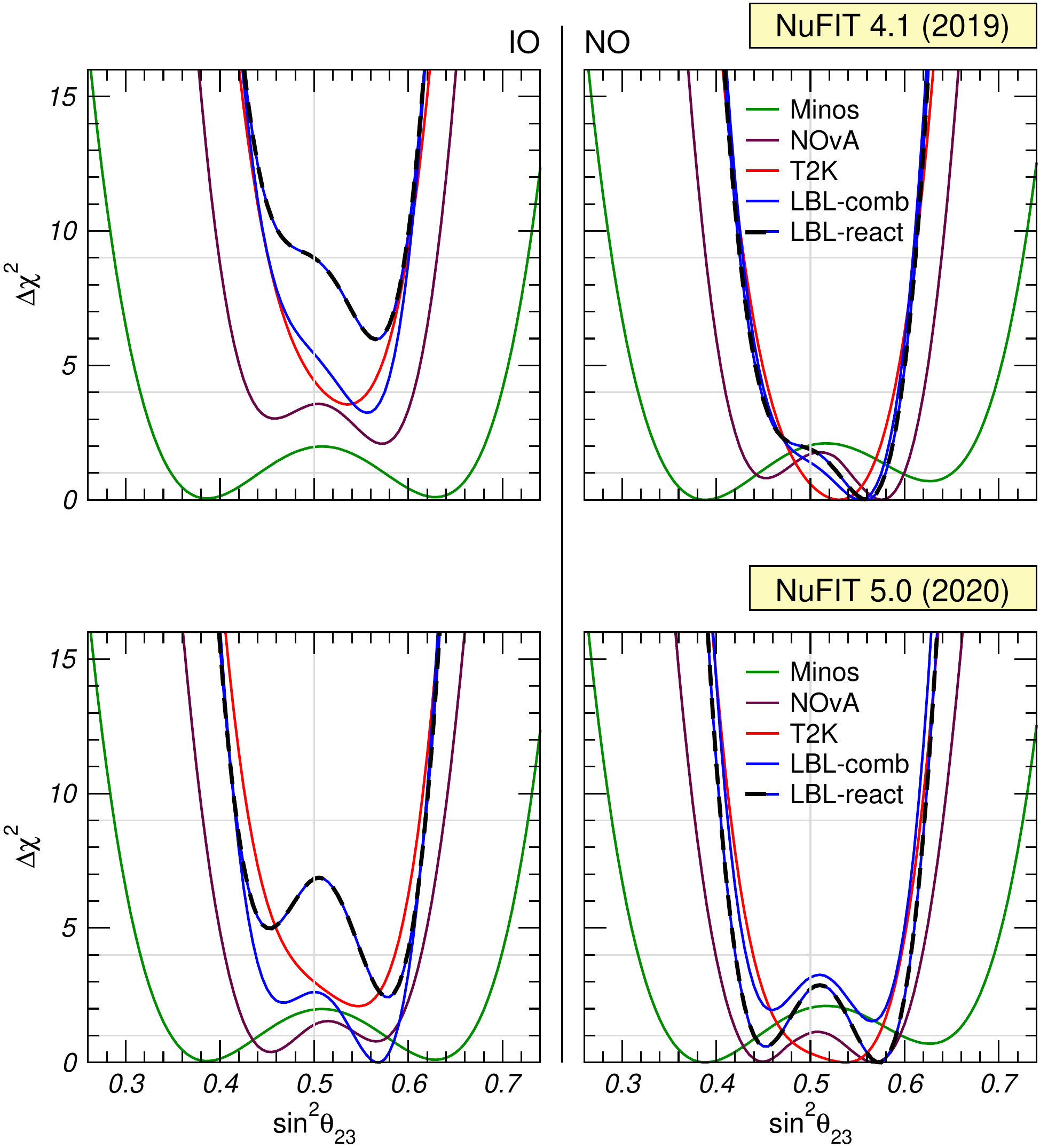}
  \caption{$\Delta\chi^2$ profiles as a function of $\Dmq_{3\ell}$
    (left) and $\sin^2\theta_{23}$ (right) for different LBL
    data sets and their combination. In the left 4 panels we show also
    the combined reactor data from Daya-Bay, RENO and Double-Chooz.
    For all curves we have fixed $\sin^2\theta_{13}=0.0224$ as well as
    the solar parameters and minimized with respect to the other
    un-displayed parameters.  $\Delta\chi^2$ is shown with respect to
    the best fit mass ordering for each curve. Upper panels are for
    the NuFIT 4.1 data set, whereas lower panels correspond to the
    current update.}
  \label{fig:compare-dma-t23}
\end{figure}

In the previous section we have discussed the status of the hints
on CP violation and
neutrino mass ordering in the latest LBL data.  In the context of
$3\nu$ mixing the relevant oscillation probabilities for the LBL
accelerator experiments depend also on $\theta_{13}$ which is most
precisely determined from reactor experiments (and on the
$\theta_{12}$ and $\Dmq_{21}$ parameters which are independently well
constrained by solar and KamLAND data).  So in our discussion, and also
to construct the $\chi^2$ curves and regions shown in
figs.~\ref{fig:compare-dcp}, ~\ref{fig:sq23-dCP}, and~\ref{fig:compare-dma-t23}
for T2K, NOvA, Minos, and the LBL-combination, those parameters
are fixed to their current best fit values.  Given the present
precision in the determination of $\theta_{13}$ this yields very
similar results to marginalize with respect to $\theta_{13}$, taking
into account the information from reactor data by adding a Gaussian
penalty term to the corresponding $\chi^2_\text{LBL}$.

Let us stress that such procedure is not the same as making a combined
analysis of LBL and reactor data, compare for instance the blue solid
versus black/blue dashed curves in fig.~\ref{fig:compare-dcp}. This is
so because relevant additional information on the mass ordering can be
obtained from the comparison of $\nu_\mu$ and $\nu_e$ disappearance
spectral data~\cite{Nunokawa:2005nx, Minakata:2006gq}. In brief, the
relevant disappearance probabilities are approximately symmetric with
respect to the sign of two effective mass-squared differences, usually
denoted as $\Dmq_{\mu\mu}$ and $\Dmq_{ee}$, respectively. They are two
different linear combinations of $\Dmq_{31}$ an $\Dmq_{32}$.
Consequently, the precise determination of the oscillation frequencies
in $\nu_\mu$ and $\nu_e$ disappearance experiments, yields information
on the sign of $\Dmq_{3\ell}$.  This effect has been present already
in previous data (see, \textit{e.g.}, Ref.~\cite{Esteban:2018azc} for
a discussion). We see from the two lower-left panels of
figure~\ref{fig:compare-dma-t23} that the region for $|\Dmq_{3\ell}|$
for IO from the LBL combination (blue curve) is somewhat in tension
with the one from the reactor experiments Daya-Bay, RENO and
Double-Chooz (black curve), while they are in quite good agreement for
NO.

In the accelerator-reactor combination this leads again to a best fit
point for NO, with $\Delta\chi^2(\text{IO}) = 2.7$, considerably less
than the value $6.2$ of NuFIT 4.1. This is explicitly shown, for example,
in the  LBL-reactor curves in fig.~\ref{fig:compare-dcp}.
For the NO best fit, a compromise
between T2K and NOvA appearance data has to be adopted, avoiding over-shoting the
number of neutrino events in NOvA while still being able to
accommodate both neutrino and anti-neutrino data from T2K, see
figure~\ref{fig:nevts}. This leads to a shift of the allowed region
towards $\dCP=\pi$ and a rather wide allowed range for $\dCP$ for NO,
see figures~\ref{fig:compare-dcp} and~\ref{fig:sq23-dCP}.  On the
other hand, we see from these figures that for IO, both T2K and NOvA
prefer $\dCP \simeq 270^\circ$. Consequently, if we restrict to this
ordering, CP conservation remains disfavored at $\sim 3\sigma$.

The behaviour as a function of $\sin^2\theta_{23}$ is shown in
fig.~\ref{fig:sq23-dCP} and the right panels of
figure~\ref{fig:compare-dma-t23}. It is mostly driven by the two T2K
neutrino samples. As follows from eq.~\eqref{eq:Nnu}, their predicted
event rate can be enhanced by increasing
$\sin^2\theta_{23}$. Therefore, in order to compensate for the
reduction in IO, a slight preference for the second $\theta_{23}$
octant emerges for IO. In case of NO, this is less preferrable, since
large $\sin^2\theta_{23}$ would worsen the T2K anti-neutrino fit as
well as NOvA neutrino data.

\subsection{Consistency between T2K, NOvA and reactors}
\label{sec:PG}
           
Let us now address the question of whether some data sets are in
tension with each other at a worrisome level. A useful method to
quantify the consistency of different data sets is the so-called
parameter goodness-of-fit (PG)~\cite{Maltoni:2003cu}. It makes use of
the following test statistic:
\begin{equation}
  \chi^2_\text{PG} = \chi^2_\text{min,glob} - \sum_i \chi^2_{\text{min}, i} \,,
\end{equation}
where $i$ labels different data sets, $\chi^2_{\text{min}, i}$ is the
$\chi^2$ minimum of each data set individually, and
$\chi^2_\text{min,glob}$ is the $\chi^2$ minimum of the global data,
\textit{i.e.}, $\chi^2_\text{min,glob} = \min\big[ \sum_i\chi^2_i
  \big]$.  Let us denote by $n_i$ the number of model parameters on
which the data set $i$ depends, and $n_\text{glob}$ the number of
parameters on which the global data depends. Then the test statistic
$\chi^2_\text{PG}$ follows a $\chi^2$ distribution with $n$ degrees of
freedom, where~\cite{Maltoni:2003cu}
\begin{equation}
  n = \sum_i n_i - n_\text{glob} \,.
\end{equation}

\begin{table}\centering
  \catcode`?=\active\def?{\hphantom{0}}
  \begin{tabular}{l|ccc|ccc}
  \hline\hline
  data sets &  \multicolumn{3}{c|}{normal ordering} 
  &  \multicolumn{3}{c}{inverted ordering} \\
  & $\chi^2_\text{PG} / n$ & $p$-value & \#$\sigma$
  & $\chi^2_\text{PG} / n$ & $p$-value & \#$\sigma$ \\
  \hline
  T2K vs NOvA          & 6.7/4 & 0.15? & $1.4\sigma$ & 3.6/4 & 0.46? & $0.7\sigma$\\
  T2K vs React         & 0.3/2 & 0.87? & $0.2\sigma$ & 2.5/2 & 0.29? & $1.1\sigma$\\  
  NOvA vs React        & 3.0/2 & 0.23? & $1.2\sigma$ & 6.2/2 & 0.045 & $2.0\sigma$ \\
  T2K vs NOvA vs React & 8.4/6 & 0.21? & $1.3\sigma$ & 8.9/6 & 0.18? & $1.3\sigma$ \\  
  \hline
  T2K vs NOvA          & 6.5/3 & 0.088 & $1.7\sigma$ & 2.8/3 & 0.42? & $0.8\sigma$ \\
  T2K vs NOvA vs React & 7.8/4 & 0.098 & $1.7\sigma$ & 7.2/4 & 0.13? & $1.5\sigma$ \\
  \hline\hline
  \end{tabular}
  \caption{Testing the consistency of different data sets shown in the
    first column assuming either normal or inverted
    ordering. ``React'' includes Daya-Bay, RENO and Double-Chooz.  In
    the analyses above the horizontal line, $\theta_{13}$ is a free
    parameter, whereas below the line we have fixed $\sin^2\theta_{13}
    = 0.0224$. See text for more details.}
    \label{tab:PG}
\end{table}
        
We are going to apply this test now to different combination of the
three data sets, ``T2K'', ``NOvA'', and ``React'', where ``React'' is
the joint data set of Daya-Bay, RENO and Double-Chooz.\footnote{We
  have also checked that the three reactor experiments by themselves
  are in excellent agreement with each other, see the figure
  ``Synergies: atmospheric mass-squared splitting'' available at
  \cite{nufit}. This justifies to merge them into a single set.} The
accelerator samples always include appearance and disappearance
channels for both neutrinos and anti-neutrinos. In order to study the
consistency of the sets under a given hypothesis for the neutrino mass
ordering, all minimizations are preformed restricting to a given mass
ordering.  Furthermore, the solar parameters are kept fixed and hence,
we have $n_\text{T2K} = n_\text{NOvA} = n_\text{glob} = 4$ (namely
$\theta_{13}$, $\theta_{23}$, $\dCP$, $|\Dmq_{3\ell}|$) and
$n_\text{React} = 2$ (namely $\theta_{13}$, $|\Dmq_{3\ell}|$).  The
results are shown in table~\ref{tab:PG}.

First, we check the pair-wise consistency of two out of the three
sets. In all cases we find perfect consistency with $p$-values well
above 10\%. The only exception is NOvA vs React for IO which show
tension at the $2\sigma$ level. A large contribution to this effect
comes from the determination of $\Dmq_{3\ell}$, which agrees better
for NO than for IO, see fig.~\ref{fig:compare-dma-t23} (lower-left
panels). The consistency of all three sets (T2K vs NOvA vs React) is
excellent for both orderings.

Second, we perform an analysis for fixed $\sin^2\theta_{13} = 0.0224$
for all data sets.  Since the accelerator experiments provide a
comparatively weak constraint on $\theta_{13}$ we want to remove this
freedom from the T2K and NOvA fits and test the consistency under the
hypothesis of fixed $\theta_{13}$. Under this assumption, all $n_i$ as
well as $n_\text{glob}$ quoted above are reduced by 1. The results of
this analysis are shown in the lower part of tab.~\ref{tab:PG}.
Testing T2K vs NOvA under this assumption, we find better
compatibility for IO, consistent with the discussion above and
figs.~\ref{fig:compare-dcp} and~\ref{fig:sq23-dCP}.  Let us stress,
however, that even for NO the $p$-value is 9\%, indicating consistency
at the $1.7\sigma$ level. Hence, we find no severe tension between T2K
and NOvA. Finally, the joint T2K vs NOvA vs React analysis with fixed
$\theta_{13}$ reveals roughly equal good consistency among the three
sets for both orderings, at around $1.5\sigma$. For NO the very slight
tension is driven by T2K vs NOvA, whereas for IO the
reactor/accelerator complementarity in the determination of
$\Dmq_{3\ell}$ provides a few units to $\chi^2_\text{PG}$.

To conclude this discussion, we find that all involved data sets are
perfectly statistically compatible under the hypothesis of
three-flavor oscillations.

\section{Resolved tension in the solar sector}
\label{sec:sol-kam}

The analyses of the solar experiments and of KamLAND give the dominant
contribution to the determination of $\Dmq_{21}$ and $\theta_{12}$.
It has been a result of global analyses for the last decade, that the
value of $\Dmq_{21}$ preferred by KamLAND was somewhat higher than the
one from solar experiments.  The tension appeared due to a combination
of two effects: the well-known fact that the \Nuc{8}{B} measurements
performed by SNO, SK and Borexino showed no evidence of the low energy
spectrum turn-up expected in the standard
LMA-MSW~\cite{Wolfenstein:1977ue, Mikheev:1986gs} solution for the
value of $\Dmq_{21}$ favored by KamLAND, and the observation of a
non-vanishing day-night asymmetry in SK, whose size is larger than the
one predicted for the $\Dmq_{21}$ value indicated by KamLAND.  In our
last published analysis~\cite{Esteban:2018azc} we included the
energy-zenith spectra or day/night spectra for SK1--3, together with
the 2860-day total energy spectrum of SK4~\cite{sksol:nu2018}.  This
last one made the lack of the turn-up effect slightly stronger.  As
for the day-night variation in SK4, it was included in terms of their
quoted day-night asymmetry for SK4 2055-day~\cite{sksol:nakano2016}
\begin{equation}
  A_\text{D/N,SK4-2055} =
  [-3.1\pm 1.6(\text{stat.}) \pm 1.4(\text{syst.})]\% \,.
\label{eq:oldadn}
\end{equation}  
Altogether this resulted in slightly over $2\sigma$ discrepancy
between the best fit $\Dmq_{21}$ value indicated of KamLAND and the
solar results.  For example the best fit $\Dmq_{21}$ of KamLAND was at
$\Delta\chi^2_\text{solar} = 4.7$ in the analysis with the GS98
fluxes.

\begin{figure}\centering
  \includegraphics[width=0.9\textwidth]{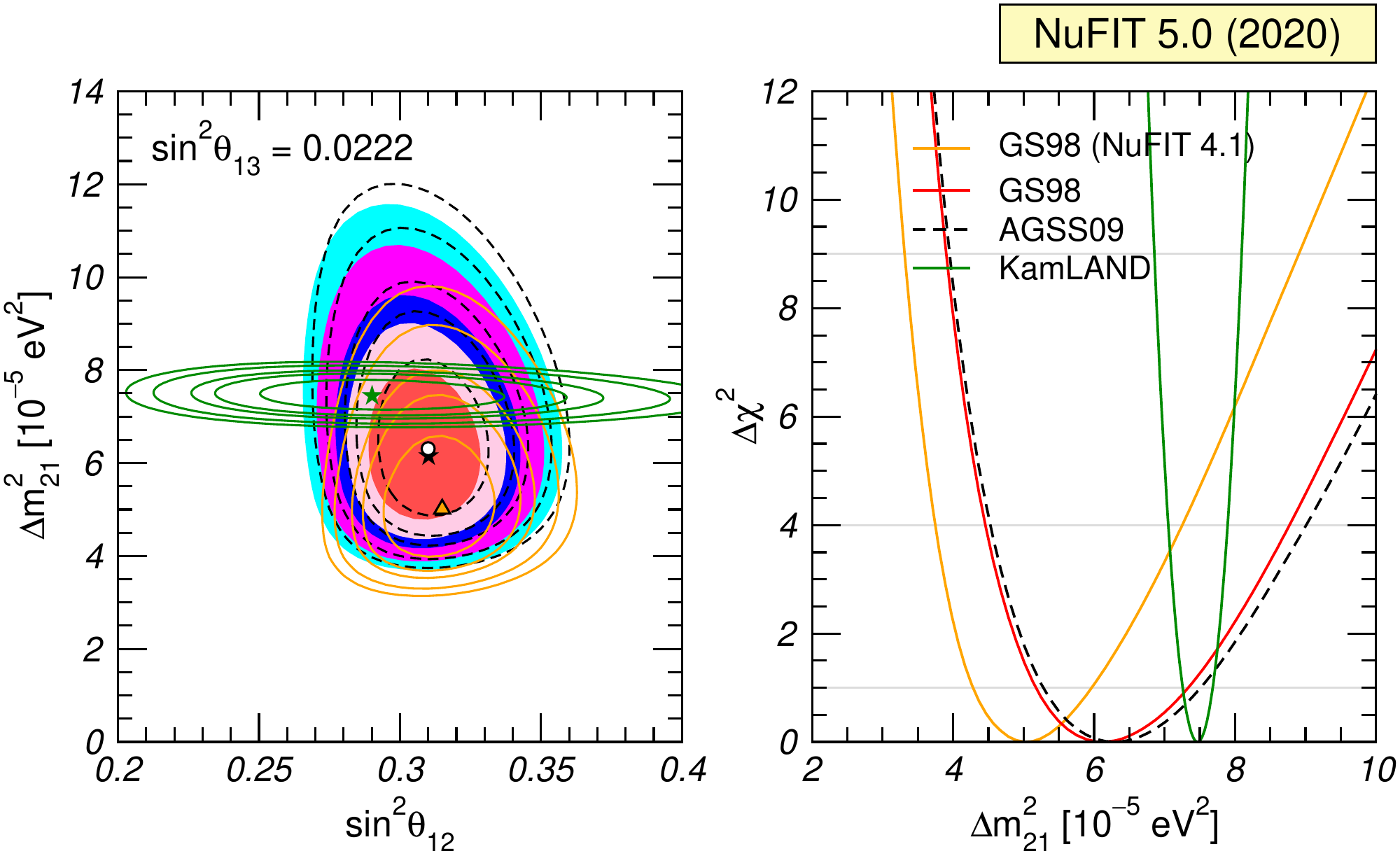}
  \caption{Left: Allowed parameter regions (at 1$\sigma$, 90\%,
    2$\sigma$, 99\%, and 3$\sigma$ CL for 2 dof) from the combined
    analysis of solar data for GS98 model (full regions with best fit
    marked by black star) and AGSS09 model (dashed void contours with
    best fit marked by a white dot), and for the analysis of KamLAND
    data (solid green contours with best fit marked by a green star)
    for fixed $\sin^2{\theta_{13}}=0.0224$ ($\theta_{13}=8.6$). We
    also show as orange contours the previous results of the global
    analysis for the GS98 model in Ref~.\cite{Esteban:2018azc}.
    Right: $\Delta\chi^2$ dependence on $\Dmq_{21}$ for the same four
    analyses after marginalizing over $\theta_{12}$.}
  \label{fig:sun-tension}
\end{figure}

Here we update the solar analysis to include the latest SK4 2970-day
results\footnote{We do not include here the latest data release from
  Borexino~\cite{Agostini:2020mfq}, which is expected to have a very
  small impact on the determination of oscillation parameters.}
presented in Neutrino2020~\cite{SK:nu2020} in the form of their total
energy spectrum and the updated day-night asymmetry
\begin{equation}
  A_\text{D/N,SK4-2970} = (-2.1\pm 1.1)\% \,.
  \label{eq:newadn}
\end{equation}  
We show in fig.~\ref{fig:sun-tension} the present determination of
these parameters from the global solar analysis in comparison with
that of KamLAND data. The results of the solar neutrino analysis are
shown for the two latest versions of the Standard Solar Model, namely
the GS98 and the AGSS09 models~\cite{Bergstrom:2016cbh} obtained with
two different determinations of the solar
abundances~\cite{Vinyoles:2016djt}. For sake of comparison we also
show the corresponding results of the solar analysis with the
pre-Neutrino2020 data~\cite{Esteban:2018azc}.

As seen in the figure, with the new data the tension between the best
fit $\Dmq_{21}$ of KamLAND and that of the solar results has
decreased. Quantitatively we now find that the best fit $\Dmq_{21}$ of
KamLAND lies at $\Delta\chi^2_\text{solar} = 1.3$ ($1.14\sigma$) in
the analysis with the GS98 fluxes.  This decrease in the tension is
due to both, the smaller day-night asymmetry (which lowers
$\Delta\chi^2_\text{solar}$ of the the best fit $\Dmq_{21}$ of KamLAND
by $2.4$ units) and the slightly more pronounced turn-up in the low
energy part of the spectrum which lowers it one extra unit.

\section{Global fit results}
\label{sec:globalsum}

Finally we present a selection of the results of our global analysis
NuFIT 5.0 using data available up to July 2020 (see
appendix~\ref{sec:appendix-data} for the complete list of the used
data including references).  We show two versions of the analysis
which differ in the inclusion of the results of the Super-Kamiokande
atmospheric neutrino data (SK-atm). As discussed in
Ref.~\cite{Esteban:2018azc} there is not enough information available
for us to make an independent analysis comparable in detail to that
performed by the collaboration, hence we have been making use of their
tabulated $\chi^2$ map which we can combine with our global analysis
for the rest of experiments.  This table was made available for their
analysis of SK1--4 corresponding to 328 kton-years
data~\cite{SKatm:data2018}.  The collaboration has presented new
oscillation results obtained from the analysis of updated SK4 samples,
both by itself~\cite{Jiang:2019xwn} and in combination with the SK1--3
phases~\cite{SK:nu2020}. They seem to indicate that their hint for
ordering discrimination has also decreased. Unfortunately the
corresponding $\chi^2$ maps of these analyses have not been made
public. Hence in what follows we refer as ``with SK-atm'' to the
analysis including the tabulated SK1--4 328 kiloton years data
$\chi^2$ map, \textit{i.e.}, the same as in NuFIT 4.0 and 4.1.

\begin{pagefigure}\centering
 \includegraphics[width=0.86\textwidth]{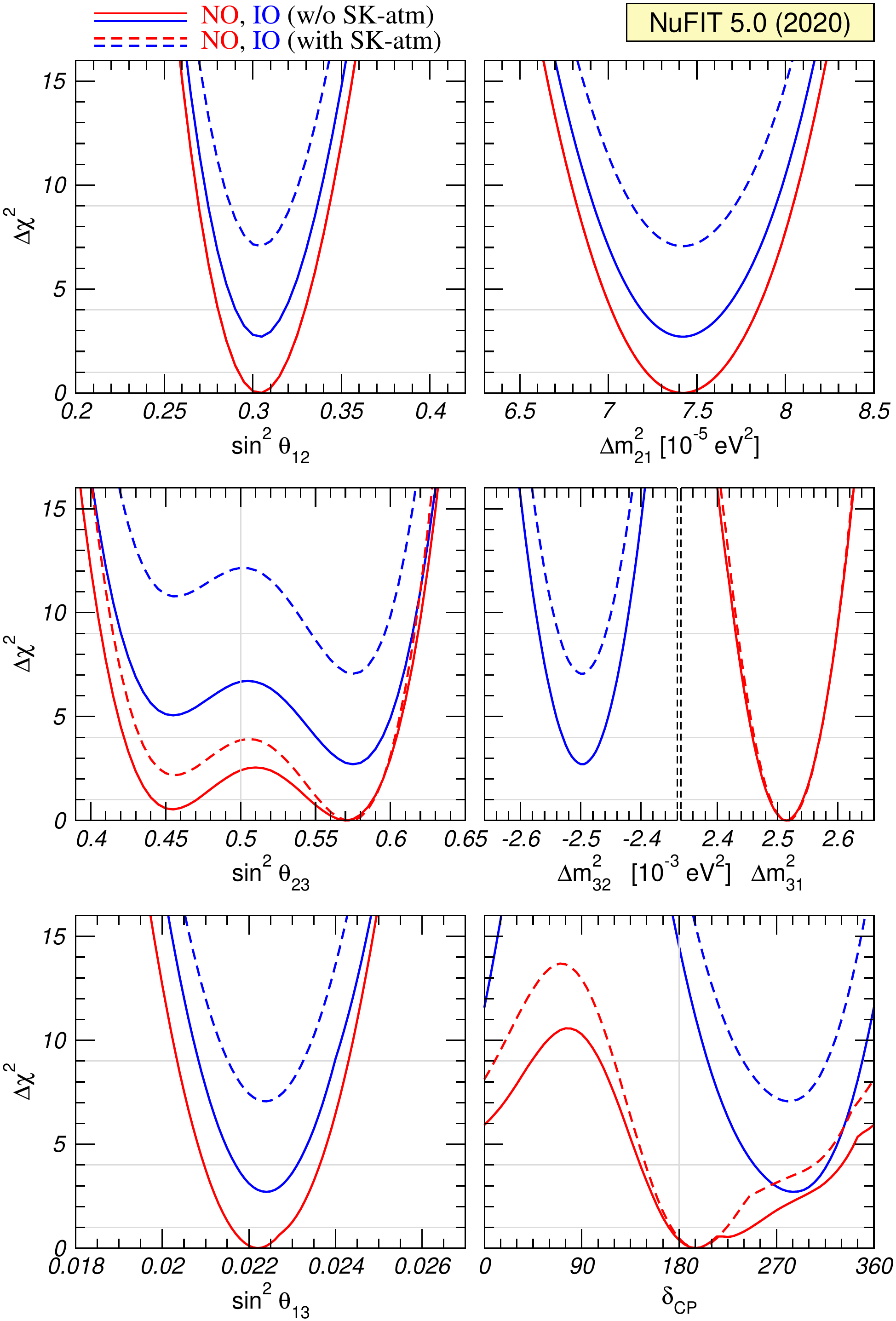}
  \caption{Global $3\nu$ oscillation analysis. We show $\Delta\chi^2$
    profiles minimized with respect to all undisplayed parameters. The
    red (blue) curves correspond to Normal (Inverted) Ordering. Solid
    (dashed) curves are without (with) adding the tabulated SK-atm
    $\Delta\chi^2$.  Note that as atmospheric mass-squared splitting
    we use $\Dmq_{31}$ for NO and $\Dmq_{32}$ for IO.}
  \label{fig:chisq-glob}
\end{pagefigure}

\begin{pagefigure}\centering
 \includegraphics[width=0.81\textwidth]{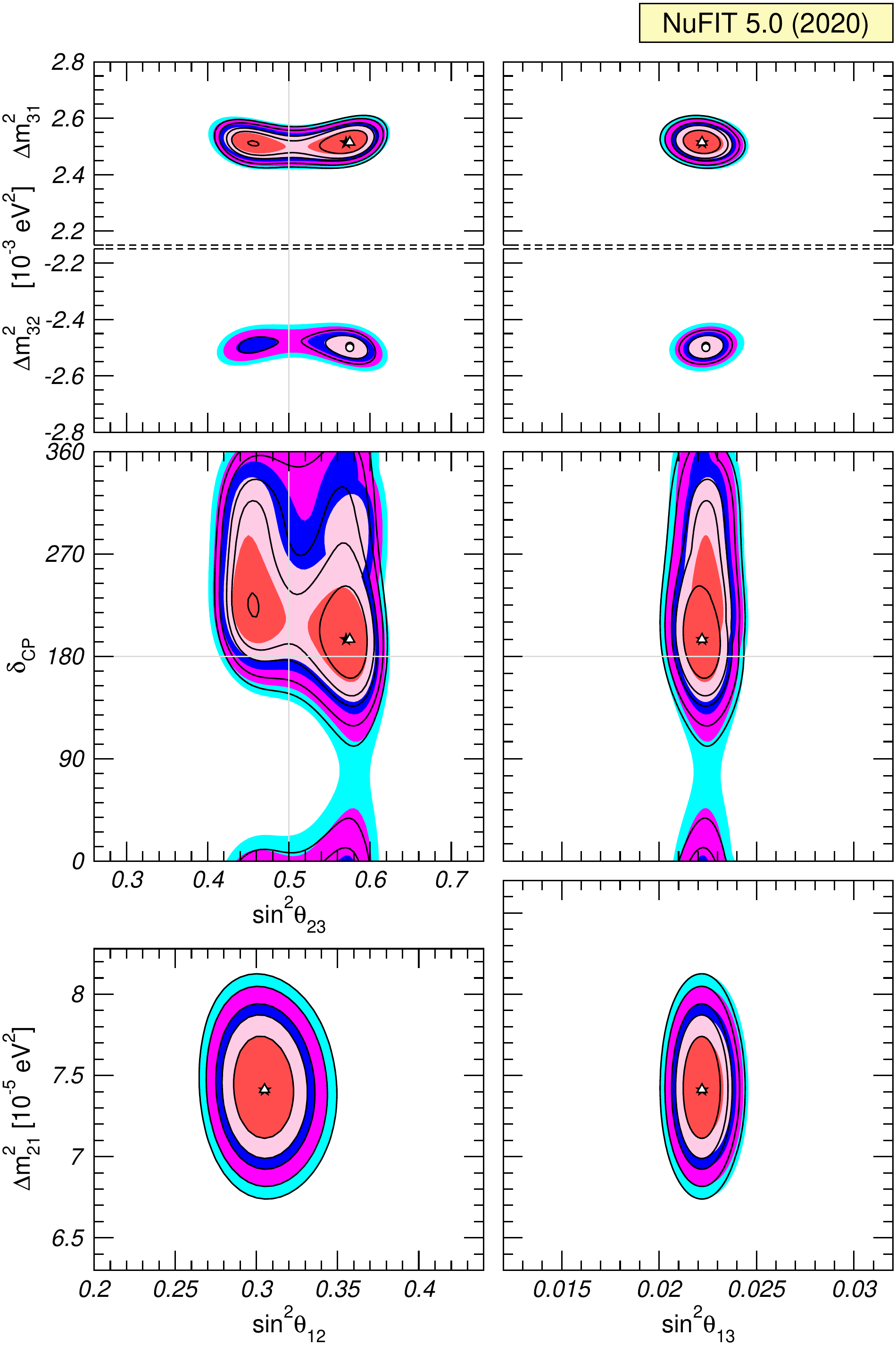}
  \caption{Global $3\nu$ oscillation analysis. Each panel shows the
    two-dimensional projection of the allowed six-dimensional region
    after minimization with respect to the undisplayed parameters. The
    regions in the four lower panels are obtained from $\Delta\chi^2$
    minimized with respect to the mass ordering. The different
    contours correspond to $1\sigma$, 90\%, $2\sigma$, 99\%, $3\sigma$
    CL (2 dof). Colored regions (black contour curves) are without
    (with) adding the tabulated SK-atm $\Delta\chi^2$. Note that as
    atmospheric mass-squared splitting we use $\Dmq_{31}$ for NO and
    $\Dmq_{32}$ for IO.}
  \label{fig:region-glob}
\end{pagefigure}

\begin{table}\centering
  \begin{footnotesize}
    \begin{tabular}{c|l|cc|cc}
      \hline\hline
      \multirow{11}{*}{\begin{sideways}\hspace*{-7em}without SK atmospheric data\end{sideways}} &
      & \multicolumn{2}{c|}{Normal Ordering (best fit)}
      & \multicolumn{2}{c}{Inverted Ordering ($\Delta\chi^2=2.7$)}
      \\
      \cline{3-6}
      && bfp $\pm 1\sigma$ & $3\sigma$ range
      & bfp $\pm 1\sigma$ & $3\sigma$ range
      \\
      \cline{2-6}
      \rule{0pt}{4mm}\ignorespaces
      & $\sin^2\theta_{12}$
      & $0.304_{-0.012}^{+0.013}$ & $0.269 \to 0.343$
      & $0.304_{-0.012}^{+0.013}$ & $0.269 \to 0.343$
      \\[1mm]
      & $\theta_{12}/^\circ$
      & $33.44_{-0.75}^{+0.78}$ & $31.27 \to 35.86$
      & $33.45_{-0.75}^{+0.78}$ & $31.27 \to 35.87$
      \\[3mm]
      & $\sin^2\theta_{23}$
      & $0.570_{-0.024}^{+0.018}$ & $0.407 \to 0.618$
      & $0.575_{-0.021}^{+0.017}$ & $0.411 \to 0.621$
      \\[1mm]
      & $\theta_{23}/^\circ$
      & $49.0_{-1.4}^{+1.1}$ & $39.6 \to 51.8$
      & $49.3_{-1.2}^{+1.0}$ & $39.9 \to 52.0$
      \\[3mm]
      & $\sin^2\theta_{13}$
      & $0.02221_{-0.00062}^{+0.00068}$ & $0.02034 \to 0.02430$
      & $0.02240_{-0.00062}^{+0.00062}$ & $0.02053 \to 0.02436$
      \\[1mm]
      & $\theta_{13}/^\circ$
      & $8.57_{-0.12}^{+0.13}$ & $8.20 \to 8.97$
      & $8.61_{-0.12}^{+0.12}$ & $8.24 \to 8.98$
      \\[3mm]
      & $\dCP/^\circ$
      & $195_{-25}^{+51}$ & $107 \to 403$
      & $286_{-32}^{+27}$ & $192 \to 360$
      \\[3mm]
      & $\dfrac{\Dmq_{21}}{10^{-5}~\eVq}$
      & $7.42_{-0.20}^{+0.21}$ & $6.82 \to 8.04$
      & $7.42_{-0.20}^{+0.21}$ & $6.82 \to 8.04$
      \\[3mm]
      & $\dfrac{\Dmq_{3\ell}}{10^{-3}~\eVq}$
      & $+2.514_{-0.027}^{+0.028}$ & $+2.431 \to +2.598$
      & $-2.497_{-0.028}^{+0.028}$ & $-2.583 \to -2.412$
      \\[2mm]
      \hline\hline
      \multirow{11}*{\begin{sideways}\hspace*{-7em}with SK atmospheric data\end{sideways}} &
      & \multicolumn{2}{c|}{Normal Ordering (best fit)}
      & \multicolumn{2}{c}{Inverted Ordering ($\Delta\chi^2=7.1$)}
      \\
      \cline{3-6}
      && bfp $\pm 1\sigma$ & $3\sigma$ range
      & bfp $\pm 1\sigma$ & $3\sigma$ range
      \\
      \cline{2-6}
      \rule{0pt}{4mm}\ignorespaces
      & $\sin^2\theta_{12}$
      & $0.304_{-0.012}^{+0.012}$ & $0.269 \to 0.343$
      & $0.304_{-0.012}^{+0.013}$ & $0.269 \to 0.343$
      \\[1mm]
      & $\theta_{12}/^\circ$
      & $33.44_{-0.74}^{+0.77}$ & $31.27 \to 35.86$
      & $33.45_{-0.75}^{+0.78}$ & $31.27 \to 35.87$
      \\[3mm]
      & $\sin^2\theta_{23}$
      & $0.573_{-0.020}^{+0.016}$ & $0.415 \to 0.616$
      & $0.575_{-0.019}^{+0.016}$ & $0.419 \to 0.617$
      \\[1mm]
      & $\theta_{23}/^\circ$
      & $49.2_{-1.2}^{+0.9}$ & $40.1 \to 51.7$
      & $49.3_{-1.1}^{+0.9}$ & $40.3 \to 51.8$
      \\[3mm]
      & $\sin^2\theta_{13}$
      & $0.02219_{-0.00063}^{+0.00062}$ & $0.02032 \to 0.02410$
      & $0.02238_{-0.00062}^{+0.00063}$ & $0.02052 \to 0.02428$
      \\[1mm]
      & $\theta_{13}/^\circ$
      & $8.57_{-0.12}^{+0.12}$ & $8.20 \to 8.93$
      & $8.60_{-0.12}^{+0.12}$ & $8.24 \to 8.96$
      \\[3mm]
      & $\dCP/^\circ$
      & $197_{-24}^{+27}$ & $120 \to 369$
      & $282_{-30}^{+26}$ & $193 \to 352$
      \\[3mm]
      & $\dfrac{\Dmq_{21}}{10^{-5}~\eVq}$
      & $7.42_{-0.20}^{+0.21}$ & $6.82 \to 8.04$
      & $7.42_{-0.20}^{+0.21}$ & $6.82 \to 8.04$
      \\[3mm]
      & $\dfrac{\Dmq_{3\ell}}{10^{-3}~\eVq}$
      & $+2.517_{-0.028}^{+0.026}$ & $+2.435 \to +2.598$
      & $-2.498_{-0.028}^{+0.028}$ & $-2.581 \to -2.414$
      \\[2mm]
      \hline\hline
    \end{tabular}
  \end{footnotesize}
  \caption{Three-flavor oscillation parameters from our fit to global
    data.  The numbers in the 1st (2nd) column are obtained assuming
    NO (IO), \textit{i.e.}, relative to the respective local minimum.
    Note that $\Dmq_{3\ell} \equiv \Dmq_{31} > 0$ for NO and
    $\Dmq_{3\ell} \equiv \Dmq_{32} < 0$ for IO. The results shown in
    the upper (lower) table are without (with) adding the tabulated
    SK-atm $\Delta\chi^2$.}
  \label{tab:bfranges}
\end{table}

Here we graphically present the results of our global analysis in the
form of one-dimensional $\Delta\chi^2$ curves
(fig.~\ref{fig:chisq-glob}) and two-dimensional projections of
confidence regions (fig.~\ref{fig:region-glob}).  The corresponding
best fit values as well as $1\sigma$ and $3\sigma$ confidence
intervals for the oscillation parameters are listed in
table~\ref{tab:bfranges}.\footnote{For additional figures and tables
  corresponding to this global analysis we refer the reader to the
  NuFIT webpage~\cite{nufit}.}  Defining the $3\sigma$ relative
precision of the parameter by $2(x^\text{up} - x^\text{low}) /
(x^\text{up} + x^\text{low})$, where $x^\text{up}$ ($x^\text{low}$) is
the upper (lower) bound on a parameter $x$ at the $3\sigma$ level, we
obtain the following $3\sigma$ relative precision (marginalizing over
ordering):
\begin{equation}
  \label{eq:precision}
  \begin{aligned}
    \theta_{12} &: 14\% \,, &\quad
    \theta_{13} &: 9.0\% \,, &\quad
    \theta_{23} &: 27\%\, [25\%] \,,
    \\
    \Dmq_{21} &: 16\% \,, &\quad
    |\Dmq_{3\ell}| &: 6.7\% \, [6.5\%] \,, &\quad
    \dCP &: 100\% \, [100\%] \,,
  \end{aligned}
\end{equation}
where the numbers between brackets show the impact of including SK-atm
in the precision of the determination of such parameter. The
$\Delta\chi^2$ profile of $\dCP$ is not gaussian and hence its
precision estimation above is only indicative.

In table~\ref{tab:bfranges} we give the best fit values and confidence
intervals for both mass orderings, relative to the local best fit
points in each ordering. The global confidence intervals
(marginalizing also over the ordering) are identical to the ones for
normal ordering, which have also been used in
eq.~\eqref{eq:precision}.  The only exception to this statement is
$\Dmq_{3\ell}$ in the analysis without SK-atm: in this case a
disconnected interval would appear above $2\sigma$ corresponding to
negative values of $\Dmq_{3\ell}$ (\textit{i.e.}, inverted ordering).

Projecting over the combinations appearing on the elements of the
leptonic mixing matrix we derive the following $3\sigma$ ranges (see
Ref.~\cite{GonzalezGarcia:2003qf} for details on how we derive the
ranges) on their magnitude:
\begin{equation}
  \label{eq:umatrix}
  \begin{aligned}
    |U|_{3\sigma}^\text{w/o SK-atm} &=
    \begin{pmatrix}
      0.801 \to 0.845 &\qquad
      0.513 \to 0.579 &\qquad
      0.143 \to 0.156
      \\
      0.233 \to 0.507 &\qquad
      0.461 \to 0.694 &\qquad
      0.631 \to 0.778
      \\
      0.261 \to 0.526 &\qquad
      0.471 \to 0.701 &\qquad
      0.611 \to 0.761
    \end{pmatrix}
    \\[1mm]
    |U|_{3\sigma}^\text{with SK-atm} &=
    \begin{pmatrix}
      0.801 \to 0.845 &\qquad
      0.513 \to 0.579 &\qquad
      0.143 \to 0.155
      \\
      0.234 \to 0.500 &\qquad
      0.471 \to 0.689 &\qquad
      0.637 \to 0.776
      \\
      0.271 \to 0.525 &\qquad
      0.477 \to 0.694 &\qquad
      0.613 \to 0.756
    \end{pmatrix}
  \end{aligned}
\end{equation}
Note that there are strong correlations between these allowed ranges
due to the unitary constraint.

\begin{figure}\centering
  \includegraphics[width=0.9\textwidth]{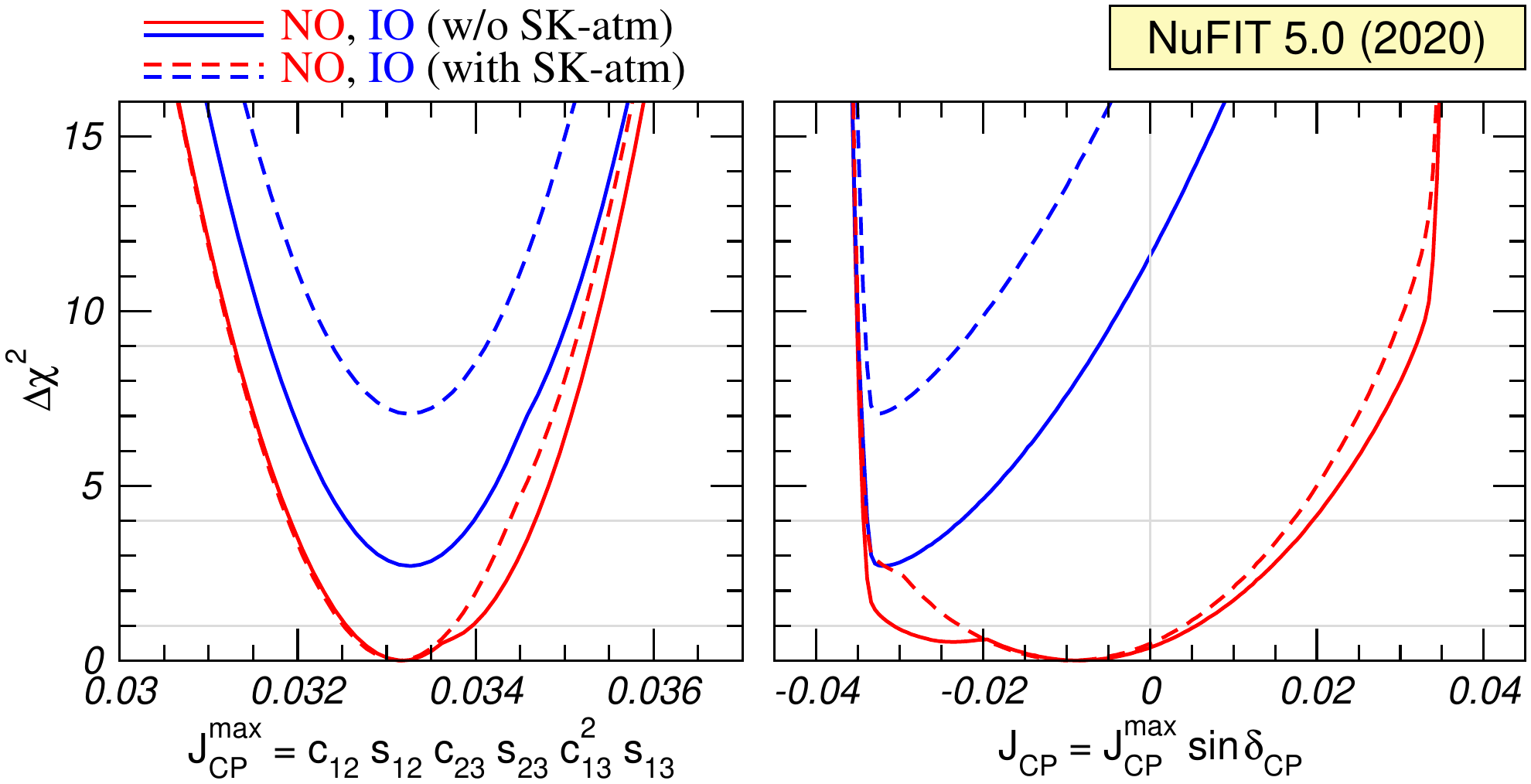}
  \caption{Dependence of the global $\Delta\chi^2$ function on the
    Jarlskog invariant. The red (blue) curves are for NO (IO). Solid
    (dashed) curves are without (with) adding the tabulated SK-atm
    $\Delta\chi^2$.}
  \label{fig:chisq-viola}
\end{figure}

The present status of leptonic CP violation is further illustrated in
fig.~\ref{fig:chisq-viola} where we show the determination of the the
Jarlskog invariant defined as:
\begin{equation}
  \begin{split}
    J_\text{CP}
    &\equiv \Im\big[ U_{\alpha i} U_{\alpha j}^* U_{\beta i}^* U_{\beta j} \big]
    \\
    &\equiv J_\text{CP}^\text{max} \sin\dCP =
    \cos\theta_{12} \sin\theta_{12}
    \cos\theta_{23} \sin\theta_{23} \cos^2\theta_{13} \sin\theta_{13}
    \sin\dCP \,.
  \end{split}
\end{equation}
It provides a convention-independent measure of leptonic CP violation
in neutrino propagation in vacuum~\cite{Krastev:1988yu} --~analogous
to the factor introduced in Ref.~\cite{Jarlskog:1985ht} for the
description of CP violating effects in the quark sector, presently
determined to be $J_\text{CP}^\text{quarks} = (3.18\pm 0.15) \times
10^{-5}$~\cite{PDG}.
From the figure we read that the determination of
the mixing angles implies a maximal possible value of the Jarlskog
invariant of
\begin{equation}
  \label{eq:jmax}
  J_\text{CP}^\text{max} = 0.0332 \pm 0.0008 \, (\pm 0.0019)
\end{equation}
at $1\sigma$ ($3\sigma$) for both orderings.  Furthermore we see that
with the inclusion of the new results, the best fit value
$J_\text{CP}^\text{best} = -0.0089$ is only favored over CP
conservation $J_\text{CP} = 0$ with $\Delta\chi^2 = 0.38$,
irrespective of SK-atm.

\section{Summary}
\label{sec:conclu}

Let us summarize the main findings resulting from the Neutrino2020
updates in neutrino oscillations.
\begin{itemize}
\item The best fit in the global analysis remains for the normal mass
  ordering, however, with reduced significance. In the global analysis
  without SK-atm, inverted ordering is disfavored only with a
  $\Delta\chi^2 = 2.7$ ($1.6\sigma$) to be compared with $\Delta\chi^2
  = 6.2$ ($2.5\sigma$) in NuFIT 4.1. This change is driven by the new
  LBL results from T2K and NOvA which indeed by themselves favor IO
  (with $\theta_{13}$ as determined by the reactor data and
  $\theta_{12}$ and $\Dmq_{21}$ by the solar and KamLAND results). The
  best fit for NO in the combined global analysis is driven by the
  better compatibility between the $\Dmq_{3\ell}$ determined in
  $\nu_\mu$ disappearance at accelerators with that from $\nu_e$
  disappearance at reactors (see left panel in
  fig.~\ref{fig:compare-dma-t23}).

\item Despite slightly different tendencies in some parameter regions,
  T2K, NOvA and reactor experiments are statistically in very good
  agreement with each other. We have performed tests of various
  experiment and analysis combinations, which all show consistency at
  a CL below $2\sigma$ (section~\ref{sec:PG}).
   
\item If atmospheric data from Super-Kamiokande is included, inverted
  ordering is disfavored with a $\Delta\chi^2 = 7.3$ ($2.7\sigma$)
  compared to $\Delta\chi^2 = 10.4$ ($3.2\sigma$) in NuFIT 4.1. Hence,
  a modest indication for NO remains. Let us note that in the recent
  Super-Kamiokande update presented at Neutrino2020~\cite{SK:nu2020}
  (with increased statistic and improved mass ordering sensitivity)
  the $\Delta\chi^2$ for IO is reduced by about 1 unit compared to the
  analysis we are using in our global fit. Therefore we expect that
  once the $\chi^2$ map for the new SK analysis becomes available, the
  combined hint in favor of NO may further decrease.
  
\item We obtain a very mild preference for the second octant of
  $\theta_{23}$, with the best fit point located at $\sin^2\theta_{23}
  = 0.57$ (slightly more non-maximal than the best fit of 0.56 in
  NuFIT 4.1), but with the local minimum in the first octant at
  $\sin^2\theta_{23} =0.455$ at a $\Delta\chi^2=0.53 \,(2.2)$ without
  (with) SK-atm. Maximal mixing ($\sin^2\theta_{23}= 0.5$)  is
  disfavored with $\Delta\chi^2 = 2.4 \, (3.9)$ without (with)
  SK-atm.

\item The best fit for the complex phase is at $\dCP = 195^\circ$.
  Compared to previous results (\textit{e.g.}, NuFIT
  4.1~\cite{nufit}), the allowed range is pushed towards the CP
  conserving value of $180^\circ$, which is now allowed at $0.6\sigma$
  with or without SK-atm. If we restrict to IO, the best fit of $\dCP$
  remains close to maximal CP violation, with CP conservation being
  disfavored at around $3\sigma$.

\item New solar neutrino data from Super-Kamiokande lead to an upward
  shift of the allowed region for $\Dmq_{21}$, which significantly
  decreased the tension between solar and KamLAND data. They are now
  compatible at $1.1\sigma$, compared to about $2.2\sigma$ for the
  pre-Neutrino2020 situation.
\end{itemize}
Overall we have witnessed decreasing significance of various
``hints'' present in previous data. This is consistent with the fate
of fluctuations which is that of fading away as time goes by.

\subsection*{Acknowledgement}

We thank Anatael Cabrera and Jonghee Yoo for correspondence conerning
the Double-Chooz and RENO analyses, respectively.  This work was
supported by the spanish grants FPA2016-76005-C2-1-P, FPA2016-78645-P,
and PID2019-105614GB-C21, by USA-NSF grant PHY-1915093, by AGAUR
(Generalitat de Catalunya) grant 2017-SGR-929.
IE acknowledges
support from the FPU program fellowship FPU15/0369.  The authors
acknowledge the support of the Spanish Agencia Estatal de
Investigacion through the grant ``IFT Centro de Excelencia Severo
Ochoa SEV-2016-0597''.

\appendix

\section{List of data used in the analysis}
\label{sec:appendix-data}

\section*{Solar experiments}

\begin{itemize}
\item \emph{External information}: Standard Solar
  Model~\cite{Vinyoles:2016djt}.

\item Chlorine total rate~\cite{Cleveland:1998nv}, 1 data point.

\item Gallex \& GNO total rates~\cite{Kaether:2010ag}, 2 data points.

\item SAGE total rate~\cite{Abdurashitov:2009tn}, 1 data point.

\item SK1 full energy and zenith spectrum~\cite{Hosaka:2005um}, 44
  data points.

\item SK2 full energy and day/night spectrum~\cite{Cravens:2008aa}, 33
  data points.

\item SK3 full energy and day/night spectrum~\cite{Abe:2010hy}, 42
  data points.

\item SK4 2970-day day-night asymmetry~\cite{SK:nu2020} and energy
  spectrum~\cite{SK:nu2020}, 24 data points.

\item SNO combined analysis~\cite{Aharmim:2011vm}, 7 data points.

\item Borexino Phase-I 741-day low-energy data~\cite{Bellini:2011rx},
  33 data points.

\item Borexino Phase-I 246-day high-energy data~\cite{Bellini:2008mr},
  6 data points.

\item Borexino Phase-II 408-day low-energy
  data~\cite{Bellini:2014uqa}, 42 data points.
\end{itemize}

\section*{Atmospheric experiments}

\begin{itemize}
\item \emph{External information}: Atmospheric neutrino
  fluxes~\cite{Honda:2015fha}.

\item IceCube/DeepCore 3-year data~\cite{Aartsen:2014yll,
  deepcore:2016}, 64 data points.

\item SK1--4 328 kiloton years~\cite{Abe:2017aap}, $\chi^2$
  map~\cite{SKatm:data2018} added to our global analysis.
\end{itemize}

\section*{Reactor experiments}

\begin{itemize}
\item KamLAND separate DS1, DS2, DS3 spectra~\cite{Gando:2013nba} with
  Daya-Bay reactor $\nu$ fluxes~\cite{An:2016srz}, 69 data points.

\item Double-Chooz FD/ND spectral ratio, with 1276-day (FD), 587-day
  (ND) exposures~\cite{DoubleC:nu2020}, 26 data points.

\item Daya-Bay 1958-day EH2/EH1 and EH3/EH1 spectral
  ratios~\cite{Adey:2018zwh}, 52 data points.

\item Reno 2908-day FD/ND spectral ratio~\cite{RENO:nu2020}, 45 data
  points.
\end{itemize}

\section*{Accelerator experiments}

\begin{itemize}
\item MINOS $10.71\times 10^{20}$~pot $\nu_\mu$-disappearance
  data~\cite{Adamson:2013whj}, 39 data points.

\item MINOS $3.36\times 10^{20}$~pot $\bar\nu_\mu$-disappearance
  data~\cite{Adamson:2013whj}, 14 data points.

\item MINOS $10.6\times 10^{20}$~pot $\nu_e$-appearance
  data~\cite{Adamson:2013ue}, 5 data points.

\item MINOS $3.3\times 10^{20}$~pot $\bar\nu_e$-appearance
  data~\cite{Adamson:2013ue}, 5 data points.

\item T2K $19.7\times 10^{20}$ pot $\nu_\mu$-disappearance
  data~\cite{T2K:nu2020}, 35 data points.

\item T2K $19.7\times 10^{20}$ pot $\nu_e$-appearance
  data~\cite{T2K:nu2020}, 23 data points for the CCQE and 16 data
  points for the CC1$\pi$ samples.

\item T2K $16.3\times 10^{20}$ pot $\bar\nu_\mu$-disappearance
  data~\cite{T2K:nu2020}, 35 data points.

\item T2K $16.3\times 10^{20}$ pot $\bar\nu_e$-appearance
  data~\cite{T2K:nu2020}, 23 data points.

\item NOvA $13.6\times 10^{20}$ pot $\nu_\mu$-disappearance
  data~\cite{NOvA:nu2020}, 76 data points.

\item NOvA $13.6\times 10^{20}$ pot $\nu_e$-appearance
  data~\cite{NOvA:nu2020}, 13 data points.

\item NOvA $12.5\times 10^{20}$ pot $\bar{\nu}_\mu$-disappearance
  data~\cite{NOvA:nu2020}, 76 data points.

\item NOvA $12.5\times 10^{20}$ pot $\bar{\nu}_e$-appearance
  data~\cite{NOvA:nu2020}, 13 data points.
\end{itemize}

\bibliographystyle{JHEP}
\bibliography{references}

\providecommand{\href}[2]{#2}\begingroup\raggedright\begin{thebibliography}{10}

\bibitem{Esteban:2016qun}
I.~Esteban, M.~C. Gonzalez-Garcia, M.~Maltoni, I.~Mart{\'\i ne}z-Soler and
  T.~Schwetz, \emph{{Updated Fit to Three Neutrino Mixing: Exploring the
  Accelerator-Reactor Complementarity}},
  \href{http://dx.doi.org/10.1007/JHEP01(2017)087}{\emph{JHEP} {\bf 01} (2017)
  087}, [\href{http://arxiv.org/abs/1611.01514}{{\tt 1611.01514}}].

\bibitem{Esteban:2018azc}
I.~Esteban, M.~C. Gonzalez-Garcia, A.~Hernandez-Cabezudo, M.~Maltoni and
  T.~Schwetz, \emph{{Global analysis of three-flavour neutrino oscillations:
  synergies and tensions in the determination of $\theta_{23}$, $\delta_{CP}$,
  and the mass ordering}},
  \href{http://dx.doi.org/10.1007/JHEP01(2019)106}{\emph{JHEP} {\bf 01} (2019)
  106}, [\href{http://arxiv.org/abs/1811.05487}{{\tt 1811.05487}}].

\bibitem{deSalas:2020pgw}
P.~de~Salas, D.~Forero, S.~Gariazzo, P.~Martinez-Mirave, O.~Mena, C.~Ternes
  et~al., \emph{{2020 Global Reassessment of the Neutrino Oscillation
  Picture}},  \href{http://arxiv.org/abs/2006.11237}{{\tt 2006.11237}}.

\bibitem{deSalas:2018bym}
P.~De~Salas, S.~Gariazzo, O.~Mena, C.~Ternes and M.~Tortola, \emph{{Neutrino
  Mass Ordering from Oscillations and Beyond: 2018 Status and Future
  Prospects}}, \href{http://dx.doi.org/10.3389/fspas.2018.00036}{\emph{Front.
  Astron. Space Sci.} {\bf 5} (2018) 36},
  [\href{http://arxiv.org/abs/1806.11051}{{\tt 1806.11051}}].

\bibitem{Capozzi:2020qhw}
F.~Capozzi, E.~Di~Valentino, E.~Lisi, A.~Marrone, A.~Melchiorri and A.~Palazzo,
  \emph{{Addendum To: Global Constraints on Absolute Neutrino Masses and Their
  Ordering}},  \href{http://arxiv.org/abs/2003.08511}{{\tt 2003.08511}}.
  [Addendum: Phys.Rev.D 101, 116013 (2020)].

\bibitem{Capozzi:2018ubv}
F.~Capozzi, E.~Lisi, A.~Marrone and A.~Palazzo, \emph{{Current Unknowns in the
  Three Neutrino Framework}},
  \href{http://dx.doi.org/10.1016/j.ppnp.2018.05.005}{\emph{Prog. Part. Nucl.
  Phys.} {\bf 102} (2018) 48--72}, [\href{http://arxiv.org/abs/1804.09678}{{\tt
  1804.09678}}].

\bibitem{Abe:2019vii}
{\scshape T2K} collaboration, K.~Abe et~al., \emph{{Constraint on the
  Matter--Antimatter Symmetry-Violating Phase in Neutrino Oscillations}},
  \href{http://dx.doi.org/10.1038/s41586-020-2177-0}{\emph{Nature} {\bf 580}
  (2020) 339--344}, [\href{http://arxiv.org/abs/1910.03887}{{\tt 1910.03887}}].
  [Erratum: Nature 583,E \textbf{16} (2020)].

\bibitem{T2K:nu2020}
P.~Dunne, ``{Latest Neutrino Oscillation Results from T2K}.'' Talk given at the
  {\it XXIX International Conference on Neutrino Physics and Astrophysics},
  Chicago, USA, June 22--July 2, 2020 (online conference).

\bibitem{Acero:2019ksn}
{\scshape NOvA} collaboration, M.~Acero et~al., \emph{{First Measurement of
  Neutrino Oscillation Parameters Using Neutrinos and Antineutrinos by Nova}},
  \href{http://dx.doi.org/10.1103/PhysRevLett.123.151803}{\emph{Phys. Rev.
  Lett.} {\bf 123} (2019) 151803}, [\href{http://arxiv.org/abs/1906.04907}{{\tt
  1906.04907}}].

\bibitem{NOvA:nu2020}
A.~Himmel, ``{New Oscillation Results from the NOvA Experiment}.'' Talk given
  at the {\it XXIX International Conference on Neutrino Physics and
  Astrophysics}, Chicago, USA, June 22--July 2, 2020 (online conference).

\bibitem{nufit}
``{NuFit webpage}.''
\newblock \href{http://www.nu-fit.org}{\tt http://www.nu-fit.org}.

\bibitem{DoubleChooz:2019qbj}
{\scshape Double Chooz} collaboration, H.~de~Kerret et~al., \emph{{First Double
  Chooz $\mathbf{\theta_{13}}$ Measurement via Total Neutron Capture
  Detection}}, \href{http://dx.doi.org/10.1038/s41567-020-0831-y}{\emph{Nature
  Phys.} {\bf 16} (2020) 558--564},
  [\href{http://arxiv.org/abs/1901.09445}{{\tt 1901.09445}}].

\bibitem{DoubleC:nu2020}
T.~Bezerra, ``{New Results from the Double Chooz Experiment}.'' Talk given at
  the {\it XXIX International Conference on Neutrino Physics and Astrophysics},
  Chicago, USA, June 22--July 2, 2020 (online conference).

\bibitem{Bak:2018ydk}
{\scshape RENO} collaboration, G.~Bak et~al., \emph{{Measurement of Reactor
  Antineutrino Oscillation Amplitude and Frequency at RENO}},
  \href{http://dx.doi.org/10.1103/PhysRevLett.121.201801}{\emph{Phys. Rev.
  Lett.} {\bf 121} (2018) 201801}, [\href{http://arxiv.org/abs/1806.00248}{{\tt
  1806.00248}}].

\bibitem{RENO:nu2020}
J.~Yoo, ``{RENO}.'' Talk given at the {\it XXIX International Conference on
  Neutrino Physics and Astrophysics}, Chicago, USA, June 22--July 2, 2020
  (online conference).

\bibitem{SK:nu2020}
Y.~Nakajima, ``{SuperKamiokande}.'' Talk given at the {\it XXIX International
  Conference on Neutrino Physics and Astrophysics}, Chicago, USA, June 22--July
  2, 2020 (online conference).

\bibitem{GonzalezGarcia:2012sz}
M.~Gonzalez-Garcia, M.~Maltoni, J.~Salvado and T.~Schwetz, \emph{{Global fit to
  three neutrino mixing: critical look at present precision}},
  \href{http://dx.doi.org/10.1007/JHEP12(2012)123}{\emph{JHEP} {\bf 1212}
  (2012) 123}, [\href{http://arxiv.org/abs/1209.3023}{{\tt 1209.3023}}].

\bibitem{Gonzalez-Garcia:2014bfa}
M.~C. Gonzalez-Garcia, M.~Maltoni and T.~Schwetz, \emph{{Updated Fit to Three
  Neutrino Mixing: Status of Leptonic CP Violation}},
  \href{http://dx.doi.org/10.1007/JHEP11(2014)052}{\emph{JHEP} {\bf 11} (2014)
  052}, [\href{http://arxiv.org/abs/1409.5439}{{\tt 1409.5439}}].

\bibitem{Kelly:2020fkv}
K.~J. Kelly, P.~A. Machado, S.~J. Parke, Y.~F. Perez~Gonzalez and
  R.~Zukanovich-Funchal, \emph{{Back to (Mass-)Squar$E_{D}$ One: the Neutrino
  Mass Ordering in Light of Recent Data}},
  \href{http://arxiv.org/abs/2007.08526}{{\tt 2007.08526}}.

\bibitem{Elevant:2015ska}
J.~Elevant and T.~Schwetz, \emph{{On the determination of the leptonic CP
  phase}}, \href{http://dx.doi.org/10.1007/JHEP09(2015)016}{\emph{JHEP} {\bf
  09} (2015) 016}, [\href{http://arxiv.org/abs/1506.07685}{{\tt 1506.07685}}].

\bibitem{Wolfenstein:1977ue}
L.~Wolfenstein, \emph{{Neutrino oscillations in matter}},
  \href{http://dx.doi.org/10.1103/PhysRevD.17.2369}{\emph{Phys. Rev.} {\bf D17}
  (1978) 2369--2374}.

\bibitem{Nunokawa:2005nx}
H.~Nunokawa, S.~J. Parke and R.~Zukanovich~Funchal, \emph{{Another Possible Way
  to Determine the Neutrino Mass Hierarchy}},
  \href{http://dx.doi.org/10.1103/PhysRevD.72.013009}{\emph{Phys. Rev.} {\bf
  D72} (2005) 013009}, [\href{http://arxiv.org/abs/hep-ph/0503283}{{\tt
  hep-ph/0503283}}].

\bibitem{Minakata:2006gq}
H.~Minakata, H.~Nunokawa, S.~J. Parke and R.~Zukanovich~Funchal,
  \emph{{Determining Neutrino Mass Hierarchy by Precision Measurements in
  Electron and Muon Neutrino Disappearance Experiments}},
  \href{http://dx.doi.org/10.1103/PhysRevD.74.053008}{\emph{Phys. Rev.} {\bf
  D74} (2006) 053008}, [\href{http://arxiv.org/abs/hep-ph/0607284}{{\tt
  hep-ph/0607284}}].

\bibitem{Maltoni:2003cu}
M.~Maltoni and T.~Schwetz, \emph{{Testing the Statistical Compatibility of
  Independent Data Sets}},
  \href{http://dx.doi.org/10.1103/PhysRevD.68.033020}{\emph{Phys. Rev.} {\bf
  D68} (2003) 033020}, [\href{http://arxiv.org/abs/hep-ph/0304176}{{\tt
  hep-ph/0304176}}].

\bibitem{Mikheev:1986gs}
S.~P. Mikheev and A.~Y. Smirnov, \emph{{Resonance enhancement of oscillations
  in matter and solar neutrino spectroscopy}}, {\emph{Sov. J. Nucl. Phys.} {\bf
  42} (1985) 913--917}.

\bibitem{sksol:nu2018}
M.~Ikeda, ``{Solar neutrino measurements with Super-Kamiokande}.'' Talk given
  at the {\it XXVIII International Conference on Neutrino Physics and
  Astrophysics}, Heidelberg, Germany, June 4--9, 2018.

\bibitem{sksol:nakano2016}
Y.~Nakano, \emph{{$^8$B solar neutrino spectrum measurement using
  Super-Kamiokande IV}}.
\newblock PhD thesis, Tokyo U., 2016-02.

\bibitem{Agostini:2020mfq}
{\scshape BOREXINO} collaboration, M.~Agostini et~al., \emph{{First Direct
  Experimental Evidence of CNO neutrinos}},
  \href{http://arxiv.org/abs/2006.15115}{{\tt 2006.15115}}.

\bibitem{Bergstrom:2016cbh}
J.~Bergstrom, M.~C. Gonzalez-Garcia, M.~Maltoni, C.~Pena-Garay, A.~M. Serenelli
  and N.~Song, \emph{{Updated determination of the solar neutrino fluxes from
  solar neutrino data}},
  \href{http://dx.doi.org/10.1007/JHEP03(2016)132}{\emph{JHEP} {\bf 03} (2016)
  132}, [\href{http://arxiv.org/abs/1601.00972}{{\tt 1601.00972}}].

\bibitem{Vinyoles:2016djt}
N.~Vinyoles, A.~M. Serenelli, F.~L. Villante, S.~Basu, J.~Bergström, M.~C.
  Gonzalez-Garcia et~al., \emph{{A new Generation of Standard Solar Models}},
  \href{http://dx.doi.org/10.3847/1538-4357/835/2/202}{\emph{Astrophys. J.}
  {\bf 835} (2017) 202}, [\href{http://arxiv.org/abs/1611.09867}{{\tt
  1611.09867}}].

\bibitem{SKatm:data2018}
{\scshape SuperKamiokande} collaboration, ``{Atmospheric neutrino oscillation
  analysis with external constraints in Super-Kamiokande I-IV}.'' link to data
  release:
  \href{http://www-sk.icrr.u-tokyo.ac.jp/sk/publications/result-e.html#atmosci2018}{\tt
  http://www-sk.icrr.u-tokyo.ac.jp/sk/publications/result-e.html\#atmosci2018},
  2018.

\bibitem{Jiang:2019xwn}
{\scshape Super-Kamiokande} collaboration, M.~Jiang et~al., \emph{{Atmospheric
  Neutrino Oscillation Analysis with Improved Event Reconstruction in
  Super-Kamiokande IV}},
  \href{http://dx.doi.org/10.1093/ptep/ptz015}{\emph{PTEP} {\bf 2019} (2019)
  053F01}, [\href{http://arxiv.org/abs/1901.03230}{{\tt 1901.03230}}].

\bibitem{GonzalezGarcia:2003qf}
M.~C. Gonzalez-Garcia and C.~Pena-Garay, \emph{{Three neutrino mixing after the
  first results from K2K and KamLAND}},
  \href{http://dx.doi.org/10.1103/PhysRevD.68.093003}{\emph{Phys. Rev.} {\bf
  D68} (2003) 093003}, [\href{http://arxiv.org/abs/hep-ph/0306001}{{\tt
  hep-ph/0306001}}].

\bibitem{Krastev:1988yu}
P.~I. Krastev and S.~T. Petcov, \emph{{Resonance Amplification and t Violation
  Effects in Three Neutrino Oscillations in the Earth}},
  \href{http://dx.doi.org/10.1016/0370-2693(88)90404-2}{\emph{Phys. Lett.} {\bf
  B205} (1988) 84--92}.

\bibitem{Jarlskog:1985ht}
C.~Jarlskog, \emph{{Commutator of the Quark Mass Matrices in the Standard
  Electroweak Model and a Measure of Maximal CP Violation}},
  \href{http://dx.doi.org/10.1103/PhysRevLett.55.1039}{\emph{Phys.Rev.Lett.}
  {\bf 55} (1985) 1039}.

\bibitem{PDG}
{\scshape Particle Data Group} collaboration, M.~Tanabashi et~al., \emph{Review
  of particle physics},
  \href{http://dx.doi.org/10.1103/PhysRevD.98.030001}{\emph{Phys. Rev. D} {\bf
  98} (Aug, 2018) 030001}.

\bibitem{Cleveland:1998nv}
B.~T. Cleveland et~al., \emph{{Measurement of the solar electron neutrino flux
  with the Homestake chlorine detector}},
  \href{http://dx.doi.org/10.1086/305343}{\emph{Astrophys. J.} {\bf 496} (1998)
  505--526}.

\bibitem{Kaether:2010ag}
F.~Kaether, W.~Hampel, G.~Heusser, J.~Kiko and T.~Kirsten, \emph{{Reanalysis of
  the GALLEX solar neutrino flux and source experiments}},
  \href{http://dx.doi.org/10.1016/j.physletb.2010.01.030}{\emph{Phys. Lett.}
  {\bf B685} (2010) 47--54}, [\href{http://arxiv.org/abs/1001.2731}{{\tt
  1001.2731}}].

\bibitem{Abdurashitov:2009tn}
{\scshape SAGE} collaboration, J.~N. Abdurashitov et~al., \emph{{Measurement of
  the solar neutrino capture rate with gallium metal. III: Results for the
  2002--2007 data-taking period}},
  \href{http://dx.doi.org/10.1103/PhysRevC.80.015807}{\emph{Phys. Rev.} {\bf
  C80} (2009) 015807}, [\href{http://arxiv.org/abs/0901.2200}{{\tt
  0901.2200}}].

\bibitem{Hosaka:2005um}
{\scshape Super-Kamiokande} collaboration, J.~Hosaka et~al., \emph{{Solar
  neutrino measurements in Super-Kamiokande-I}},
  \href{http://dx.doi.org/10.1103/PhysRevD.73.112001}{\emph{Phys. Rev.} {\bf
  D73} (2006) 112001}, [\href{http://arxiv.org/abs/hep-ex/0508053}{{\tt
  hep-ex/0508053}}].

\bibitem{Cravens:2008aa}
{\scshape Super-Kamiokande} collaboration, J.~Cravens et~al., \emph{{Solar
  neutrino measurements in Super-Kamiokande-II}},
  \href{http://dx.doi.org/10.1103/PhysRevD.78.032002}{\emph{Phys. Rev.} {\bf
  D78} (2008) 032002}, [\href{http://arxiv.org/abs/0803.4312}{{\tt
  0803.4312}}].

\bibitem{Abe:2010hy}
{\scshape Super-Kamiokande} collaboration, K.~Abe et~al., \emph{{Solar neutrino
  results in Super-Kamiokande-III}},
  \href{http://dx.doi.org/10.1103/PhysRevD.83.052010}{\emph{Phys. Rev.} {\bf
  D83} (2011) 052010}, [\href{http://arxiv.org/abs/1010.0118}{{\tt
  1010.0118}}].

\bibitem{Aharmim:2011vm}
{\scshape SNO} collaboration, B.~Aharmim et~al., \emph{{Combined Analysis of
  All Three Phases of Solar Neutrino Data from the Sudbury Neutrino
  Observatory}},
  \href{http://dx.doi.org/10.1103/PhysRevC.88.025501}{\emph{Phys. Rev.} {\bf
  C88} (2013) 025501}, [\href{http://arxiv.org/abs/1109.0763}{{\tt
  1109.0763}}].

\bibitem{Bellini:2011rx}
{\scshape Borexino} collaboration, G.~Bellini et~al., \emph{{Precision
  measurement of the 7Be solar neutrino interaction rate in Borexino}},
  \href{http://dx.doi.org/10.1103/PhysRevLett.107.141302}{\emph{Phys. Rev.
  Lett.} {\bf 107} (2011) 141302}, [\href{http://arxiv.org/abs/1104.1816}{{\tt
  1104.1816}}].

\bibitem{Bellini:2008mr}
{\scshape Borexino} collaboration, G.~Bellini et~al., \emph{{Measurement of the
  solar 8B neutrino rate with a liquid scintillator target and 3 MeV energy
  threshold in the Borexino detector}},
  \href{http://dx.doi.org/10.1103/PhysRevD.82.033006}{\emph{Phys. Rev.} {\bf
  D82} (2010) 033006}, [\href{http://arxiv.org/abs/0808.2868}{{\tt
  0808.2868}}].

\bibitem{Bellini:2014uqa}
{\scshape BOREXINO} collaboration, G.~Bellini et~al., \emph{{Neutrinos from the
  primary proton–proton fusion process in the Sun}},
  \href{http://dx.doi.org/10.1038/nature13702}{\emph{Nature} {\bf 512} (2014)
  383--386}.

\bibitem{Honda:2015fha}
M.~Honda, M.~Sajjad~Athar, T.~Kajita, K.~Kasahara and S.~Midorikawa,
  \emph{{Atmospheric Neutrino Flux Calculation Using the Nrlmsise-00
  Atmospheric Model}},
  \href{http://dx.doi.org/10.1103/PhysRevD.92.023004}{\emph{Phys. Rev.} {\bf
  D92} (2015) 023004}, [\href{http://arxiv.org/abs/1502.03916}{{\tt
  1502.03916}}].

\bibitem{Aartsen:2014yll}
{\scshape IceCube} collaboration, M.~Aartsen et~al., \emph{{Determining
  neutrino oscillation parameters from atmospheric muon neutrino disappearance
  with three years of IceCube DeepCore data}},
  \href{http://dx.doi.org/10.1103/PhysRevD.91.072004}{\emph{Phys. Rev.} {\bf
  D91} (2015) 072004}, [\href{http://arxiv.org/abs/1410.7227}{{\tt
  1410.7227}}].

\bibitem{deepcore:2016}
{\scshape IceCube} collaboration, J.~P. Yanez et~al., ``{IceCube Oscillations:
  3 years muon neutrino disappearance data}.''
  \href{http://icecube.wisc.edu/science/data/nu_osc}{\tt
  http://icecube.wisc.edu/science/data/nu\_osc}.

\bibitem{Abe:2017aap}
{\scshape Super-Kamiokande} collaboration, K.~Abe et~al., \emph{{Atmospheric
  Neutrino Oscillation Analysis with External Constraints in Super-Kamiokande
  I-IV}}, \href{http://dx.doi.org/10.1103/PhysRevD.97.072001}{\emph{Phys. Rev.}
  {\bf D97} (2018) 072001}, [\href{http://arxiv.org/abs/1710.09126}{{\tt
  1710.09126}}].

\bibitem{Gando:2013nba}
{\scshape KamLAND} collaboration, A.~Gando et~al., \emph{{Reactor On-Off
  Antineutrino Measurement with Kamland}},
  \href{http://dx.doi.org/10.1103/PhysRevD.88.033001}{\emph{Phys. Rev.} {\bf
  D88} (2013) 033001}, [\href{http://arxiv.org/abs/1303.4667}{{\tt
  1303.4667}}].

\bibitem{An:2016srz}
{\scshape Daya Bay} collaboration, F.~P. An et~al., \emph{{Improved Measurement
  of the Reactor Antineutrino Flux and Spectrum at Daya Bay}},
  \href{http://dx.doi.org/10.1088/1674-1137/41/1/013002}{\emph{Chin. Phys.}
  {\bf C41} (2017) 013002}, [\href{http://arxiv.org/abs/1607.05378}{{\tt
  1607.05378}}].

\bibitem{Adey:2018zwh}
{\scshape Daya Bay} collaboration, D.~Adey et~al., \emph{{Measurement of
  Electron Antineutrino Oscillation with 1958 Days of Operation at Daya Bay}},
  \href{http://dx.doi.org/10.1103/PhysRevLett.121.241805}{\emph{Phys. Rev.
  Lett.} {\bf 121} (2018) 241805}, [\href{http://arxiv.org/abs/1809.02261}{{\tt
  1809.02261}}].

\bibitem{Adamson:2013whj}
{\scshape MINOS} collaboration, P.~Adamson et~al., \emph{{Measurement of
  Neutrino and Antineutrino Oscillations Using Beam and Atmospheric Data in
  MINOS}}, \href{http://dx.doi.org/10.1103/PhysRevLett.110.251801}{\emph{Phys.
  Rev. Lett.} {\bf 110} (2013) 251801},
  [\href{http://arxiv.org/abs/1304.6335}{{\tt 1304.6335}}].

\bibitem{Adamson:2013ue}
{\scshape MINOS} collaboration, P.~Adamson et~al., \emph{{Electron neutrino and
  antineutrino appearance in the full MINOS data sample}},
  \href{http://dx.doi.org/10.1103/PhysRevLett.110.171801}{\emph{Phys. Rev.
  Lett.} (2013) }, [\href{http://arxiv.org/abs/1301.4581}{{\tt 1301.4581}}].

\end{thebibliography}\endgroup

\end{document}